\documentclass[12pt,preprint]{aastex}

\shorttitle{TASS Mark IV Survey}
\shortauthors{Droege, Richmond, Sallman and Creager}

\begin{document}


\title{TASS Mark IV Photometric Survey of the Northern Sky}


\author{Thomas F. Droege}
\affil{The Amateur Sky Survey}
\email{droege@tass-survey.org}

\author{Michael W. Richmond}
\affil{Physics Department, Rochester Institute of Technology, Rochester, NY 14623-5603}
\email{mwrsps@rit.edu}

\author{Michael P. Sallman}
\affil{The Amateur Sky Survey}
\email{msallman@pro-ns.net}

\author{Robert P. Creager}
\affil{The Amateur Sky Survey}
\email{Robert\_Creager@LogicalChaos.org}




\begin{abstract}
The Amateur Sky Survey (TASS) is a loose confederation of amateur and
professional astronomers. We describe the design and construction of our Mark
IV systems, a set of wide-field telescopes with CCD cameras
which take simultaneous images in the $V$ and $I_C$ passbands.
We explain our observational procedures
and the pipeline which processes and reduces the images
into lists of stellar positions and magnitudes.
We have compiled a large database of measurements 
for stars 
in the northern celestial hemisphere
with $V$-band magnitudes in the range 
$7 < V < 13$.
This paper describes data taken over the four-year
period starting November, 2001.
One of our
results is a catalog of repeated measurements on the 
Johnson-Cousins system for over 4.3 million stars.
\end{abstract}





\section{Introduction}

In recent years, the combination of relatively inexpensive
CCD detectors and powerful desktop computers 
has made it possible for almost anyone to acquire and process
very large amounts of astronomical information.
Certain projects requiring large
amounts of telescope time are
impractical at heavily oversubscribed
observatories,
but 
can now be carried out by dedicated amateurs
\citep{Paczynski2000}.
Several years ago,
we used modest equipment to perform a photometric survey 
of bright sources near the celestial equator 
\citep{Richmond2000}.
The simple, fixed mounts and drift-scan technique
of our Mark III systems
limited that project to a small strip of the 
sky.

We have since constructed new mounts
and switched to stare-mode observations,
allowing us to cover nearly the entire northern
celestial hemisphere.
We report here on results from our
Mark IV systems, 
which provide repeated $V$-band and $I_C$-band
measurements of stars of intermediate
brightness.
Section~\ref{hardwaresection}
describes our equipment,
section~\ref{operationsection}
our location and mode of operation,
and section~\ref{softwaresection}
the pipeline which extracts measurements
from our images and performs
preliminary calibration.
After discovering systematic errors in our
photometry, 
we attempted to remove their effects
from some of the derived quantities
in our database;
section~\ref{patchessection}
explains these corrections.
The resulting ``patches catalog''
provides a 
homogeneous network
averaging 190 stars per square degree
with $V$-band and $I_C$-band
measurements 
of moderate precision:
$\sim 0.05$ mag at the bright end,
$\sim 0.20$ mag at the faint end.

Our survey may be most useful
as a source of quick photometric
calibration:
it is dense enough to provide at least
one star within the small field of 
view of typical CCD images,
and its $I_C$-band measurements 
extend into the red end of the the visible portion of the spectrum,
where some unfiltered systems are most
sensitive.
Our work is, of course, no replacement
or substitute for primary photometric standards
such as 
\cite{Landolt1992},
but may serve as a temporary measure
if conditions do not permit the
necessary all-sky observations.
Studies of long-period, high-amplitude
variable stars may also profit from 
our repeated measurements over several 
years.

\section{Hardware\label{hardwaresection}}

We describe here the detectors, optics, and mounts of
our instruments; readers can skip to 
Table \ref{hardware_summary} for a summary.
The Mark IV units are built around CCD chips designated
Loral Fairchild CCD442A, 
which are $2048\times2048$ arrays of 15-micron pixels.
We designed and built our own electronics
to read the sensors;
they provide 16 bits of data per pixel
with a readout time of 46 seconds.
The gain of the CCDs is about 2.4 electrons per 
Analog-to-Data Unit (ADU),
and the readnoise is about 15 electrons.
We circulate water from a commercial chiller
through thermoelectric coolers
in each camera to keep the chips at a
temperature fixed within 1 degree Celsius throughout each night;
we vary this temperature depending on the ambient conditions,
but it is typically $-20$ degrees C.
The dark current at this temperature
is about 0.2 electrons per second per pixel,
which is negligible compared to contributions
from the background sky.
The CCDs are not isolated by a vacuum;
instead, we pump dry air through the camera heads 
to prevent moisture from condensing into ice crystals
on the silicon.

\begin{figure}
\plotone{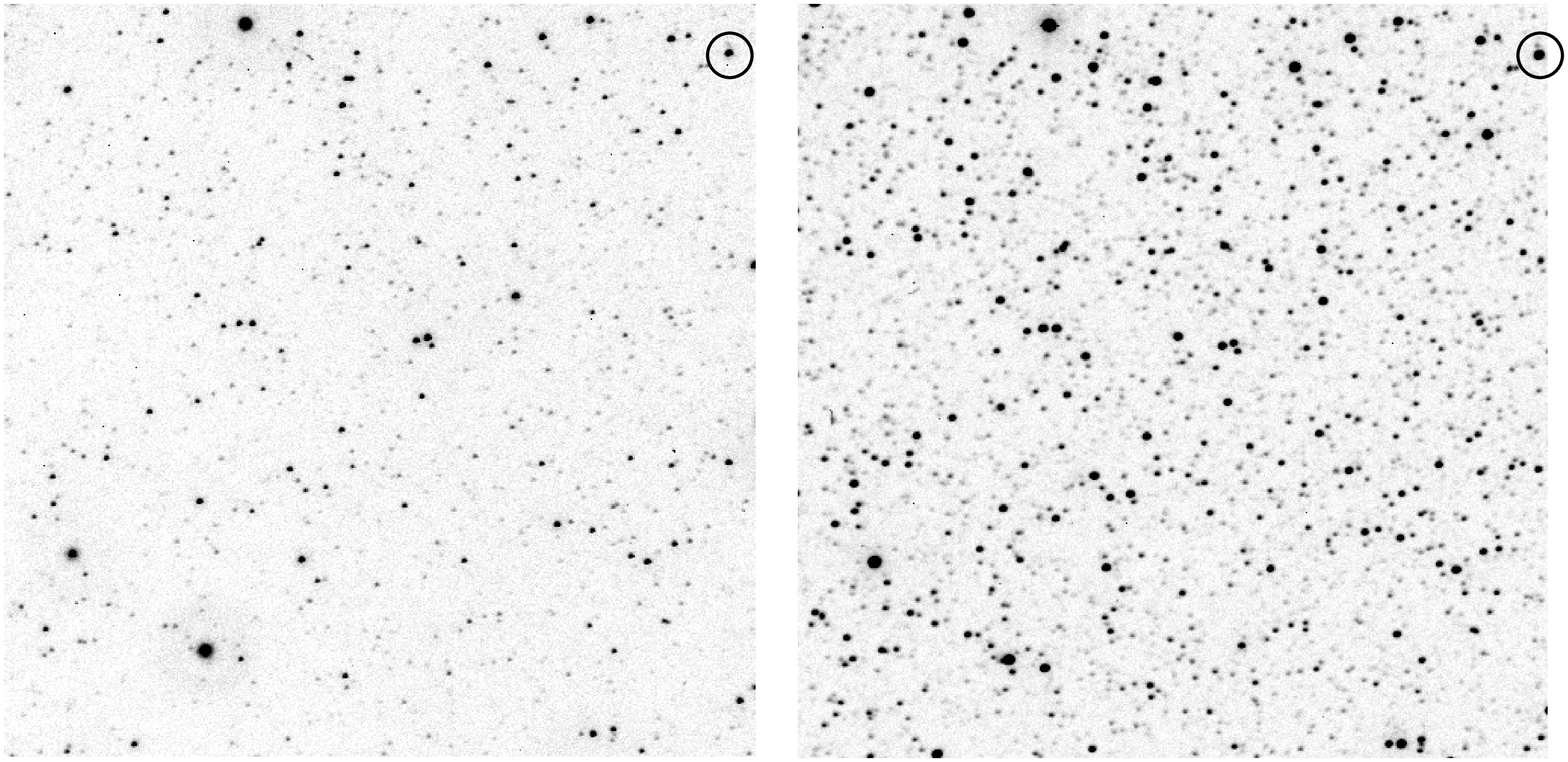}
\caption{The central portion
           of a Mark IV image pair,
           a $1^{\circ} \times 1^{\circ}$ area centered at
           RA=20:16:14.2, Dec=+16:16:18,
           about 10 degrees from the galactic plane.
           $V$-band is on the left, $I_C$-band on the right.
           North is up, East to the left.
           The circled star is R Sge.\label{centerimage}}
\end{figure}

\begin{figure}
\epsscale{1.00}
\plotone{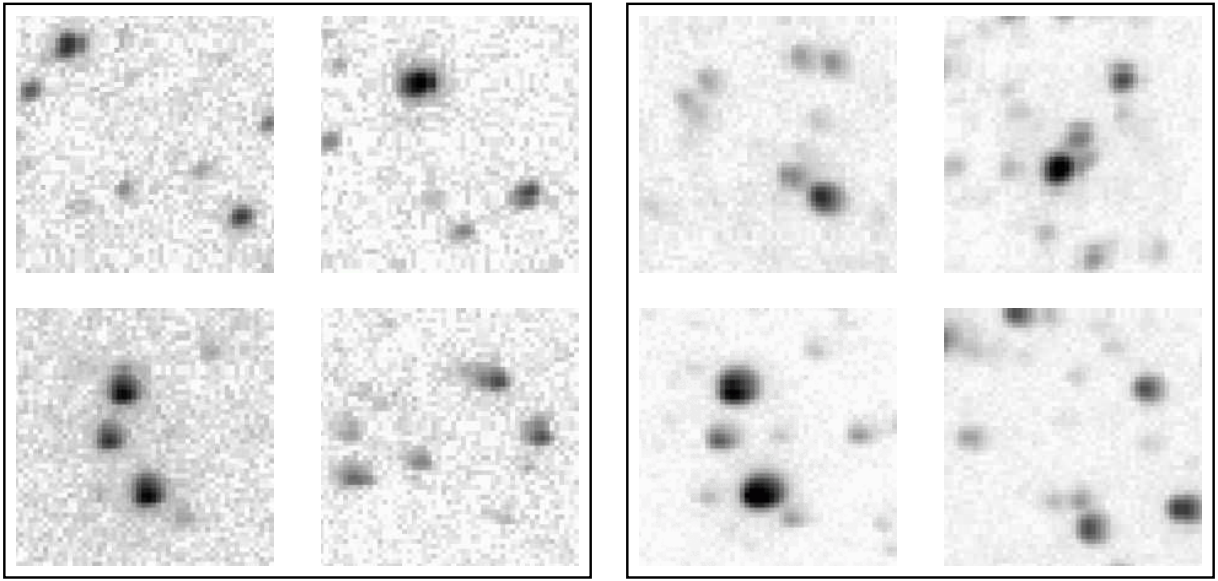}
\caption{Closeups of the corners
           of a Mark IV image pair.
           The contrast is logarithmic to enhance
           features in the PSF.
           $V$-band is on the left, $I_C$-band on the right.
           North is up, East to the left.
           \label{cornerimage}}
\end{figure}

Each CCD is mounted at the focus of its own 
small telescope.
We designed 100mm f/4 custom refractive optics to provide
small aberrations over a wide field
for a particular range of wavelengths.
%
%
%
Each lens has coatings optimized for its
portion of the spectrum.
The optics yield an image scale of 7.7 arcseconds
per pixel 
and a field about 4.2 degrees on a side.
The illumination of the focal plane 
is uniform to about $10\%$ across the width of the CCDs.
Stellar images are slightly sharper at the
center of the frame than near the edges,
with a FWHM of $2.5-3.3$ pixels in $V$-band
and $2.8-3.8$ pixels in $I_C$-band.
This corresponds to $19-29$ arcseconds,
far larger than the diffraction limit
($\sim 3^{''}$) 
or the local seeing
($\sim 3-6^{''}$).
A portion of this excess blurring is due to
the drive mechanism:
many images show a slight elongation in the
East-West direction,
and the residuals of Mark IV positions
from the Tycho-2 catalog
\citep{Hog2000} 
are slightly larger in the East-West 
direction than the North-South direction.
The remainder of the PSF width
is probably due to mediocre focusing:
we do not adjust the focus as the 
temperature changes.
The wide PSF does provide one benefit:
it avoids the photometric complications
of undersampled images
\citep{Bakos2004}.
Figure~\ref{centerimage} shows the central portion
of one pair of images,
in a relatively dense star field,
and figure~\ref{cornerimage} shows 
closeup subsections from each corner;
note the slight elongation of the
PSF away from the optical axis.

Each camera has its own filter mounted in front
of the CCD.
The filters were made by Omega Optical, Inc., according
to the 
\citet{Bessell1990}
prescription.
The shutter sits between the filter
and the lens assembly;
it has two sliding leaves which meet at the middle
of the field and take 0.2 seconds to slide together
or apart.

A single Mark IV unit consists of two telescopes ---
one $V$-band, one $I_C$-band ---
placed side-by-side on a common mount.
The equatorial mount can cover Declinations
ranging from the horizon to the pole,
but has a limited range in Right Ascension:
it can move only thirty degrees away from the meridian
in either direction.

\section{Operation\label{operationsection}}

All the data described in this paper was collected
from the roof
of the first author's home in Batavia, IL:
latitude $41.8271$ North, longitude $88.3125$ West,
altitude $217$ meters.
This region of North America does not have very 
clear skies in general,
as long experience at Yerkes Observatory
\citep[e.g.,][]{Cudworth1985}
and other nearby sites has shown;
moreover, the very bright lights of the 
Chicago metropolitan area are less than 50 km distant,
and a large interstate highway lies 6 km to the south.
We decided to acquire measurements as frequently
as possible, and therefore observed at times
through thin clouds and high humidity, and during
all lunar phases.
The bright glow and strong gradients in sky brightness
are responsible for some of the systematic 
photometric errors described in 
section~\ref{patchessection}.

Three Mark IV units contributed measurements, each 
focusing on a different region of the sky.
The unit ``TOM1'' acquired images centered 
on Declinations $-4 < \delta < +16$
over the period 2001 Nov 11 to 2005 Nov 11;
``TOM2'' scanned the region
$+20 < \delta < +48$ over the period 2003 Apr 1 to 2005 Nov 17;
and 
``TOM3'' covered the area
$+52 < \delta < +88$ between 2003 Apr 9 and 2005 Nov 17.
\footnote{At the time of writing, we have shifted the 
    zones of each unit slightly: the central positions 
    of the strips for each camera now run 
    $-4 < \delta < +20$ for TOM1, $+24 < \delta < +52$
    for TOM2, and $+56 < \delta < +88$ for TOM3.}
Each scan overlaps its neighbors to the South and North
by about one quarter of a degree, so that the three
units working together cover the entire Declination
range seamlessly.

At the start of each night, 
the operator opens the telescope covers,
points each unit to the meridian,
and sets the Declination of each
unit to the bottom of its range
(i.e., TOM1 points to $\delta = -4$).
The units then follow a strict pattern
during the night:
they take a set of images at fixed Right Ascension
while moving upwards in Declination by steps
of $4$ degrees.
Exposure times were $100$ seconds
from 2001 to 2004,
and $200$ seconds during 2005.
The systems require
about $60$ seconds
to read each set of images and slew
to the next position.
After reaching the northern limit of
its region,
each unit returns to the southern limit
and returns to the meridian.
The result is a set of slightly overlapping images
which sample the sky as it crosses the meridian.
Each unit acquires about 20 ($V$, $I_C$) image pairs per hour,
which correspond to about 660 MBytes of raw data.

Over the course of many nights,
therefore, 
the survey images occur at the same Declinations
(centered at $\delta = -4, 0, +4, +8, ...$).
However, the starting position in Right Ascension
was not set according to any rule prior to 2005,
and thus the Right Ascension centers of those
survey images vary randomly from one
night to the next.
As a consequence, 
stars in those  images
appear at roughly the same
image coordinate in one direction (from North to South),
but cover the entire range of image coordinates
in the other direction (from East to West).
We did fix the Right Ascension centers starting in 2005.

In the morning, the operator closes the telescope
covers and stows the units.
He then visually inspects five to ten percent 
of the images taken during the night, as well
as the results of the pipeline processing which
has finished so far.\footnote{
The pipeline software completes its analysis of
each image several hours after it is acquired.}
The software will find and flag some problems, such as 
periods of clouds during the night,
but other problems, such as ice crystals forming
on the CCD, are better found by a human.
If there is evidence for very poor sky conditions or equipment
problems, the entire night's data is discarded.
Individual images with obvious problems are also
deleted at this point.
Otherwise, the operator waits for the 
analysis software to complete its task,
then backs up the results.

After several months of experience, we learned that
we can control the temperature of the CCDs well enough
that dark frames need not be taken every night;
we find that a new set every month is sufficient.
We also noticed that our flatfield frames,
made from a median of many target images on a good
night, varied very little from night to night;
therefore, we update our flatfield frames
only when necessary:
that means whenever we make adjustments to the equipment,
or when the visible appearance of the raw images
changes noticeably (e.g., due to shifting dust specks),
or about once per month by default.

\section{Software\label{softwaresection}}

We have written our own software to reduce the
raw images into lists of stars with measured properties.
Some of it has grown from the
PCVista package
\citep{Treffers1989},
some of it was created specifically for this survey.
All the software is available freely online;\footnote{
  http://spiff.rit.edu/tass/pipeline/}
contact the second author for details.
We believe that one particular module may be 
of use to other astronomers, so we describe
it at some length in the Appendix.

The reduction procedure involves
four basic steps: creating master darks and flats,
applying them to target images,
finding stars and measuring their instrumental properties,
and converting the instrumental units to standard ones.
We describe each step briefly below.

To create master dark frames, we take a series of at least
ten images of the same length as our target images,
with the dome open and the cameras near the middle of their
declination range, 
but with the camera shutters closed.
The temperature of the CCD is controlled to be
the same during these dark frames as it is during normal
observations.
We compute the interquartile mean at each pixel location
and place it into the master dark frame.
To create master flatfield frames, we start with a series
of 38 target images taken during a good night.
The bright skies in Batavia provide 3000 to 10000 photoelectrons
per pixel in the background of each image, 
so the statistical variations in the combined signal
of each pixel are less than $0.3\%$.
We use a strip of prescan columns to check for any 
fixed offset between these target frames and the master dark image;
if a difference exists, we subtract it from all pixels in
the target images.  We then subtract the master dark frame
from each target image,
and then compute the pixel-by-pixel interquartile mean 
from the set of all target images to serve as the master flatfield frame.
We scan each master flatfield frame for small regions of connected
pixels which lie far from the local mean value,
due to chip defects, dust or ice crystals,
and create a mask of all such bad regions.
We later flag any detected objects which touch these bad regions.

In order to clean a raw target image,
we first compare its prescan columns to the master dark
frame's prescan columns; 
if a difference exists, we shift the target image's 
pixels by the mean difference.
We then subtract the master dark frame from the target image.
We divide the resulting target frame by a normalized version
of the master flatfield image.
We fit a gaussian to the histogram of pixel values in the 
cleaned image to estimate the sky level;
images with sky values outside a particular range 
are discarded.
We remove large-scale variations in the sky background
(due to clouds or lights in neighboring houses)
by fitting a first-order polynomial to local sky
measurements on a $10 \times 10$ grid and subtracting
the model from the image.

We search for stars in a cleaned image by marking as
candidates all pixels exceeding a fixed threshold above
the sky and measuring their properties;
candidates which pass a series of tests 
based on Full-Width at Half Maximum (FWHM),
sharpness and roundness 
\citep{Stetson1987}
are designated as stars.
We then measure each star's position on the image
by fitting one-dimensional gaussians to the intensity-weighted
marginal sums of pixels in each direction.
We calculate the instrumental magnitude of each
star using a fixed circular aperture of radius $4$ pixels
($30^{''}$)
and local sky measured as the median of values in
an annulus of radii $10$ and $20$ pixels
($77^{''}$ and $154^{''}$).
The uncertainty in this magnitude is estimated using
the statistics of electrons within the aperture
\citep{Howell1989}.
We set flags for any measurement which is likely to be
unreliable, if it is close to the edge of an image,
touches a bad region, contains saturated pixels, etc.

In order to transform the 
$(x, y)$ pixel positions of objects into
$(\rm{RA}, \rm{Dec})$,
we create a reference catalog by selecting 
roughly 
80\% of the stars from the
Tycho-2 catalog
\citep{Hog2000},
those 
which meet the ``astrometry'' criteria 
listed in 
table~\ref{tychotable}.
There are typically 50 such stars per image.
For each image, we use the 
{\tt match} 
software
(see Appendix)
to match the reference stars to 
detected objects,
fit a cubic model to the transformation
between image $(x, y)$ and 
projected $(\rm{RA}, \rm{Dec})$ coordinates,
and apply that model to the positions
of all detected objects.
The residuals from the model range from
$\sim 0{\rlap.}^{''}8$ 
for bright ($V < 10$) stars to 
$\sim 3{\rlap.}^{''}5$ for faint ($V > 13$) stars.
Our matching routines fail in regions 
very close to the celestial pole, 
forcing us to discard measurements 
with Declinations above $+88\rlap{.}^{\circ}2$.

We apply a preliminary photometric calibration
to measurements in the final stage of the pipeline.
We again choose Tycho-2 as our reference, but
create a second subset with more stringent limits
(see table~\ref{tychotable});
this yields roughly 10 to 15 reference
stars in each Mark IV image.
We convert the Tycho-2 measurements in
$B_T$ and $V_T$ to
the Johnson-Cousins $V$ and $I_C$ magnitudes
using relationships kindly provided by
Arne Henden 
\citep{Henden2001}
and shown in
table~\ref{convert_mags}.
We discard any stars not detected 
simultaneously in both $V$ and $I$.
We then create photometric solutions
for each night with the following form:
\begin{eqnarray}
V & = & v + a_j + b*(v - i) - k_V*X \\
I_C & = & i + c_j + d*(v - i) - k_I*X 
\end{eqnarray}
where
$V$ and $I_C$ are the calibrated magnitudes of a star,
$v$ and $i$ the instrumental measurements,
$a_j$ and $c_j$ the zero points of the $j$-th images during the night,
$b$ and $d$ color terms,
$k_V$ and $k_I$ the first-order extinction coefficients,
and
$X$ the airmass of each star.
Note that we allow the zero-points to vary from one 
image to the next, but assume single color terms for
the entire night.
Differential extinction can be significant over a single
frame, which may span a range of up to 0.12 airmasses,
but our limited range of 
observations does not allow us to solve for it 
reliably; we therefore assume fixed values of
$k_V = 0.20$ and $k_I = 0.06$.
If the actual extinction coefficients were twice as
large as our adopted values, 
the maximum error across a single frame would be
about $0.020$ mag in $V$-band and 
about $0.007$ mag in $I_C$-band
for objects near the southern and northern limits
of our survey.

The output of the pipeline is a list of 
stars detected simultaneously by both cameras
of a single Mark IV unit,
with (RA, Dec) positions and preliminary ($V$, $I_C$) magnitudes.
A good night yields several thousand stars per image
and (near the equinoxes) about 400 images.
At intervals of roughly one month,
we place these measurements into
the
Mark IV engineering database,\footnote{
\url{http://sallman.tass-survey.org/} }
where they are freely available.
The software is robust enough that it can run to
completion even during mediocre observing conditions,
causing a small amount of poor quality data
to enter the database.
We warn the potential user not to accept
blindly the results of every query;
checking individual measurements against those of nearby
stars and those from other nights
can help to identify bad data.
We also 
note that the preliminary magnitudes stored in
this database suffer from small systematic errors,
which we describe in the following section.

\section{The ``patches'' catalog \label{patchessection} }

\begin{figure}
\plotone{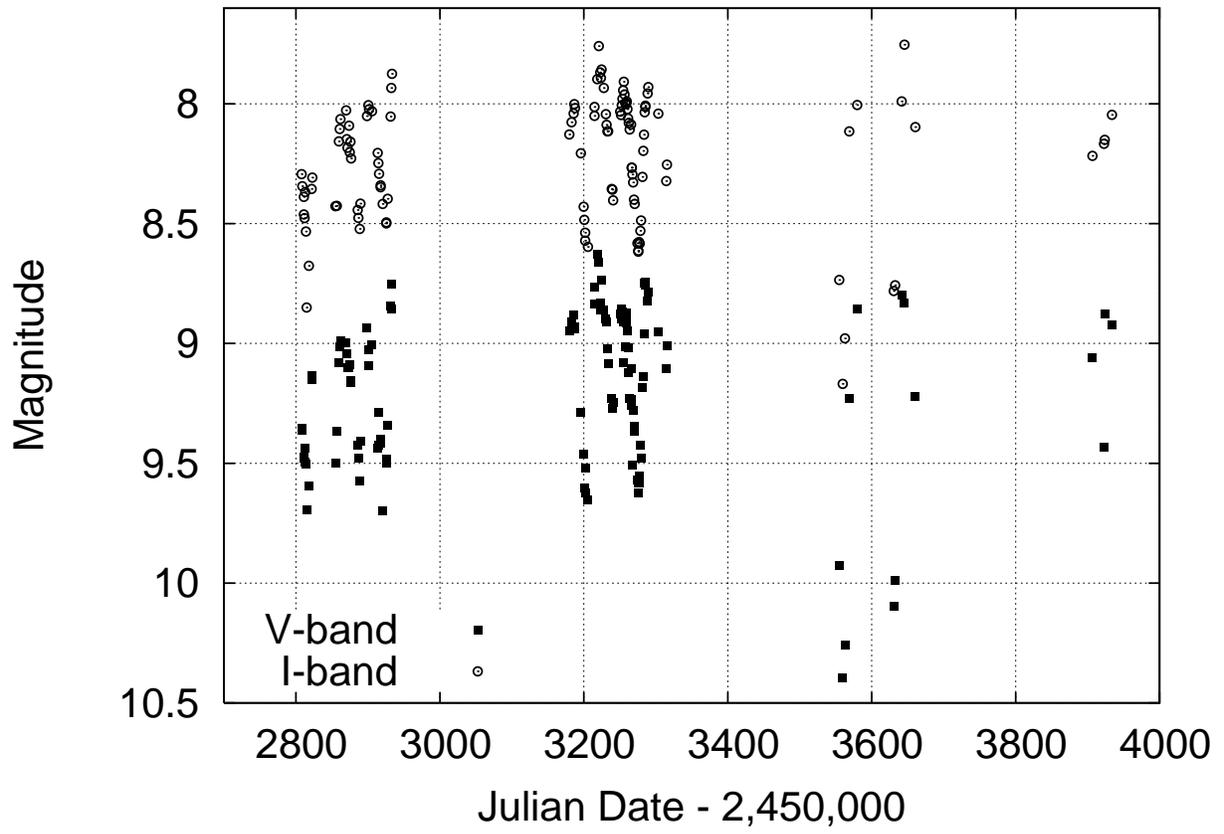}
\caption{Light curve of R Sge, using all measurements in 
           the Mark IV engineering database.\label{lightcurve}}
\end{figure}

For some purposes, the measurements produced by
the pipeline and stuffed into the engineering database
are good enough;
in
figure~\ref{lightcurve},
one can see clearly the light curve of the long-term,
large-amplitude variable star R Sge.
For others, however, it is necessary to improve
the photometric calibration of the Mark IV results.
We describe here a subset of the Mark IV data 
which we subject to additional analysis;
the end result will be a catalog of mean properties
of stars which were measured many times.
We believe that this ``patches catalog''
will be more useful to most members of the community
than the individual measurements in the engineering database.

\subsection{Selecting stars for further analysis}

\begin{figure}
\plotone{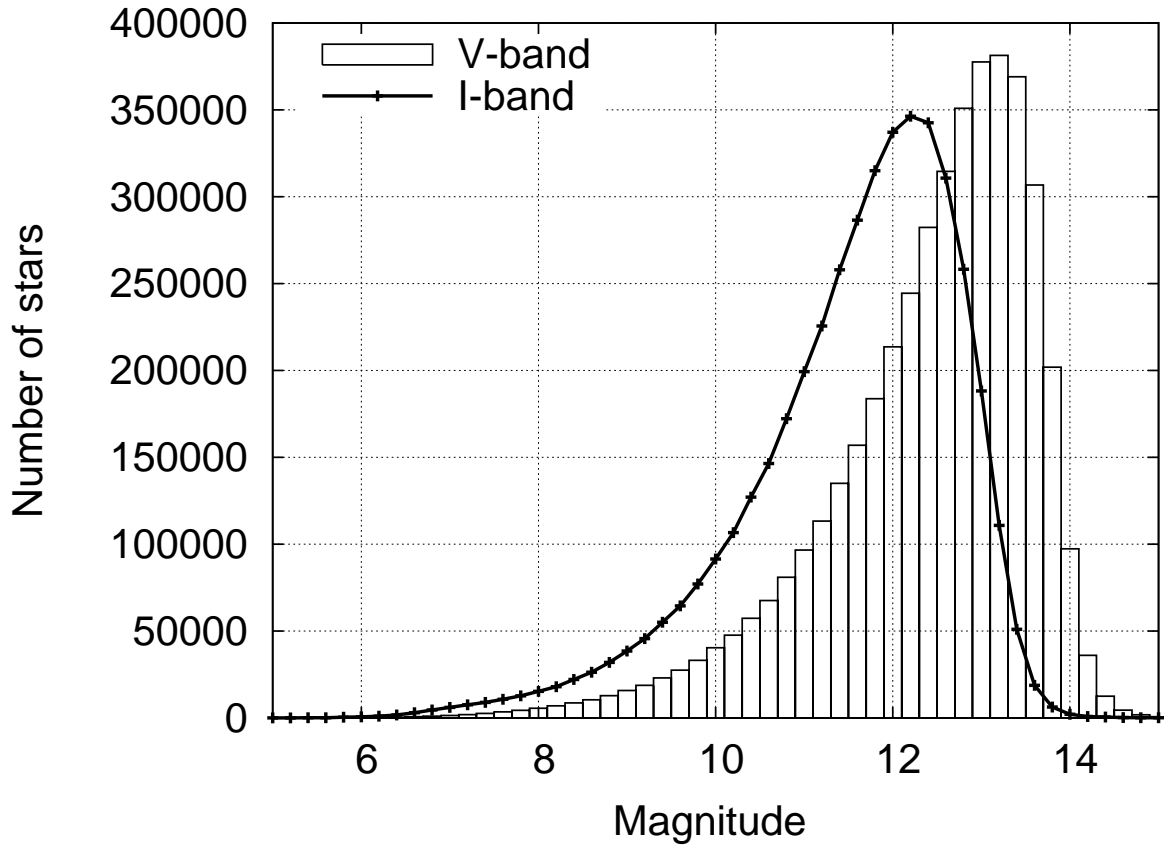}
\caption{Distribution of magnitudes for stars in the
          ``patches'' dataset.\label{maghist}}
\end{figure}

\begin{figure}
\plotone{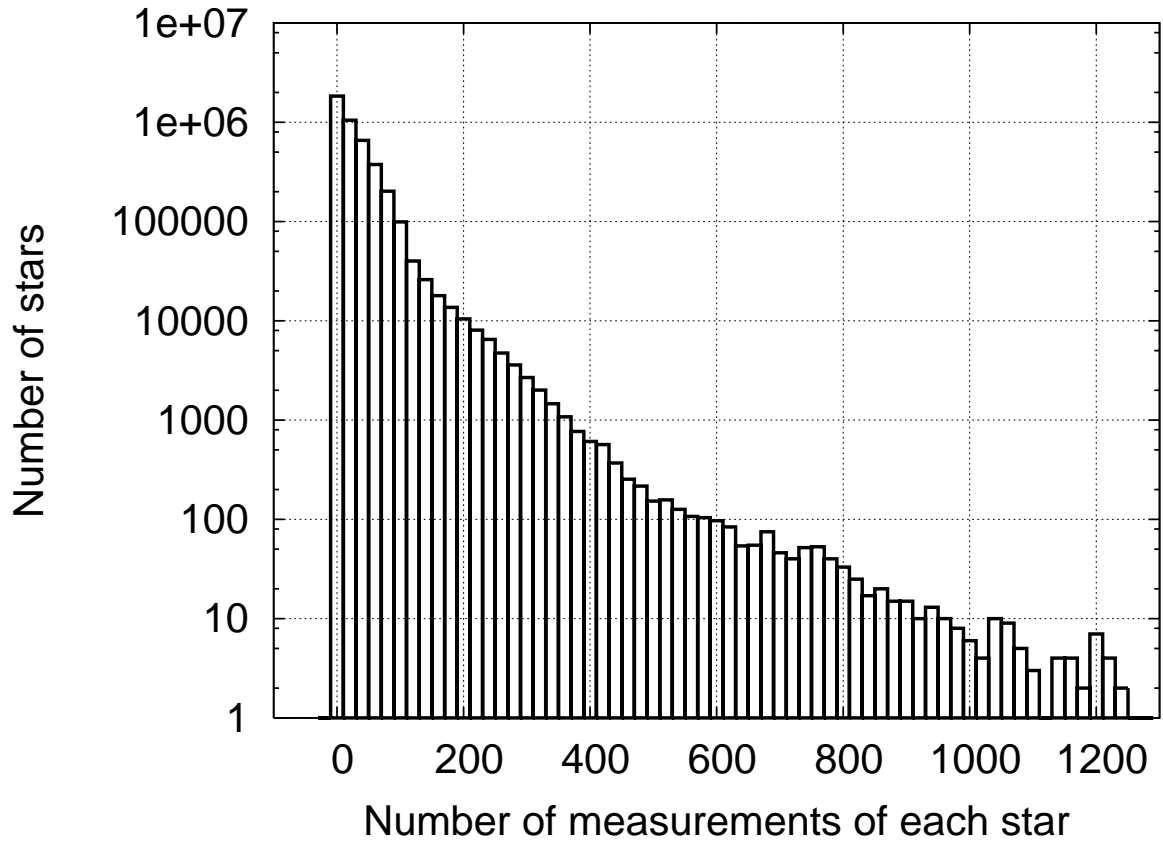}
\caption{Distribution of number of observations for stars in the 
          ``patches'' dataset.\label{numobspatches}}
\end{figure}

We began with all measurements made over the period
2001 Nov 11 to 2005 Nov 17.
Our first step was to remove spurious and unreliable
detections by selecting objects which were detected
many times.\footnote{
When new measurements are added to the database,
they are compared to the mean positions of existing
stars.
If a detection falls within $7\rlap.^{''}2$ of a
star, it is assigned to the star
and the star's mean position is updated;
otherwise, a new star entry is created in the database
and the detection assigned as its first measurement.}
We chose stars detected in at least 5
pairs of
$V, I$ images
(though we had to decrease this threshold to 3 detections 
in three isolated 
degree-sized regions of the sky with poor coverage).
The result is a set of roughly 4.3 million stars,
with magnitudes roughly in the range
$7 < V < 13$ (figure~\ref{maghist}).
The mean number of measurement pairs for each
star is 37, but the distribution has a long
tail; see figure~\ref{numobspatches}.

During our analysis of this dataset,
we found frequent outliers in the Mark IV 
measurements, due to clouds,
cosmic rays, satellite trails, 
passing airplanes, and other image defects.
In order to reduce their influence on the 
majority of the measurements, we will frequently employ an
interquartile mean (IQM) to determine 
average values.
We form the IQM by sorting all magnitude
measurements, discarding the top 25\% and bottom 25\%
of the distribution, and computing the mean of the
remaining values.

\subsection{Additional photometric calibration}

In order to check the accuracy of our photometry,
we matched the stars in the ``patch'' dataset
against stars measured by 
Landolt 
\citep{Landolt1983,Landolt1992}.
Because the Mark IV cameras have pixels over $7^{''}$
in size
and we measure light in apertures of radius $30^{''}$,
our measurements blend together light from stars
within roughly one arcminute of each other;
for comparison, Landolt made his measurements
through considerably smaller apertures, 
$13^{''}$ and $7^{''}$ in radius.
We used the 
Vizier\footnote{
\url{http://vizier.u-strasbg.fr/viz-bin/VizieR}
}
facility to check each Landolt star 
for significant neighbors,
which we define as stars from the USNO B1.0 catalog
{\citep{Monet2003}}
within $70^{''}$
and $2$ magnitudes of the Landolt star.
Stars with significant neighbors were discarded.
We also checked each Landolt star for 
variability using the GCVS version 4.2
{\citep{Samus2004}};
we discarded any star even suspected of being variable.
We examined the Mark IV record for each remaining
Landolt star and discarded any objects with significant
variability.
There are 153 isolated, constant Landolt stars
which match items in the ``patches'' dataset;
these are shown as small symbols in the figures below.
Some of these are so bright that they saturate the
Mark IV detectors, or so faint that their
measurements have very large scatter.
For some purposes, we will consider only
the 99 stars 
in the range
$8.0 < V < 12.5$ and $7.5 < I_C < 12.5$,
which have accurate and reasonably precise Mark IV photometry;
we denote them with large symbols.

\begin{figure}
\plotone{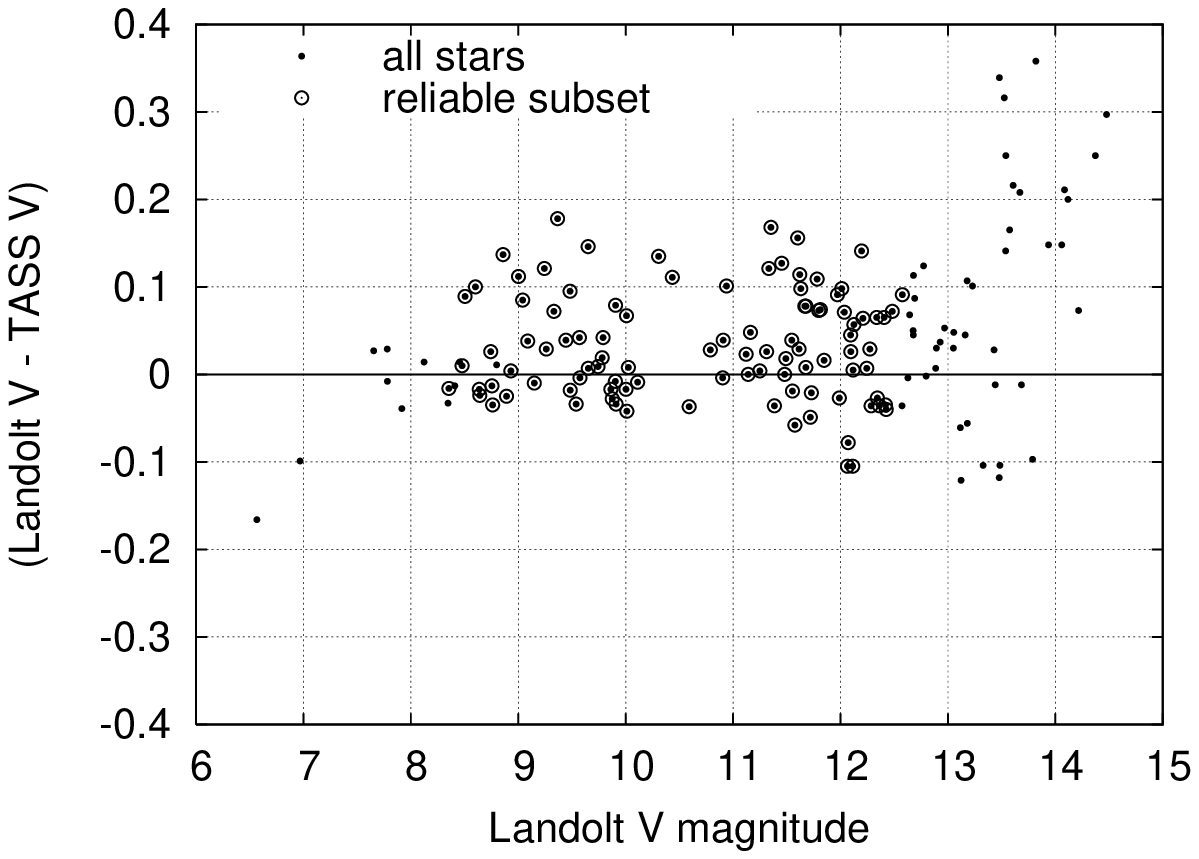}
\caption{Differences in $V$-band between Landolt photometry and interquartile 
           mean values from the Mark IV engineering database
           \label{rawvmagdiff}.}
\end{figure}

\begin{figure}
\plotone{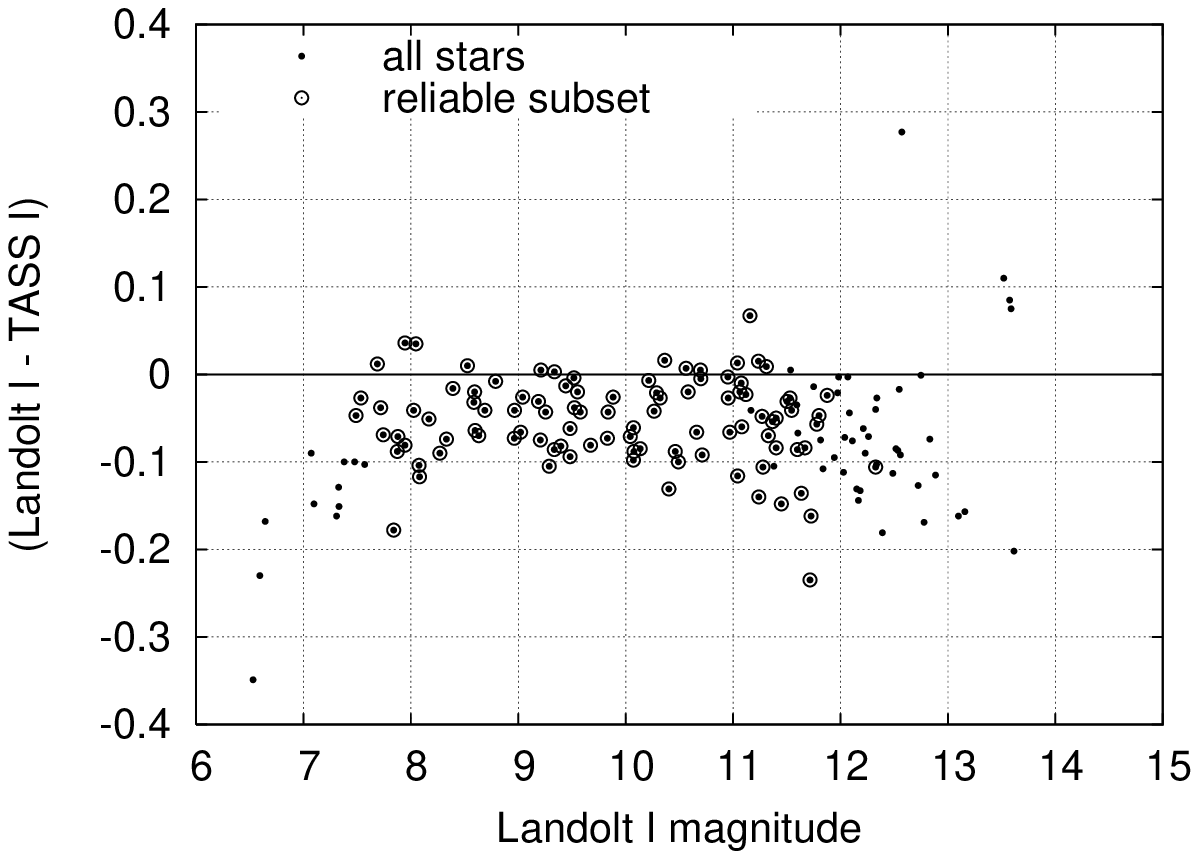}
\caption{Differences in $I_C$-band between Landolt photometry and interquartile
           mean values from the Mark IV engineering database
           \label{rawimagdiff}.}
\end{figure}

\begin{figure}
\plotone{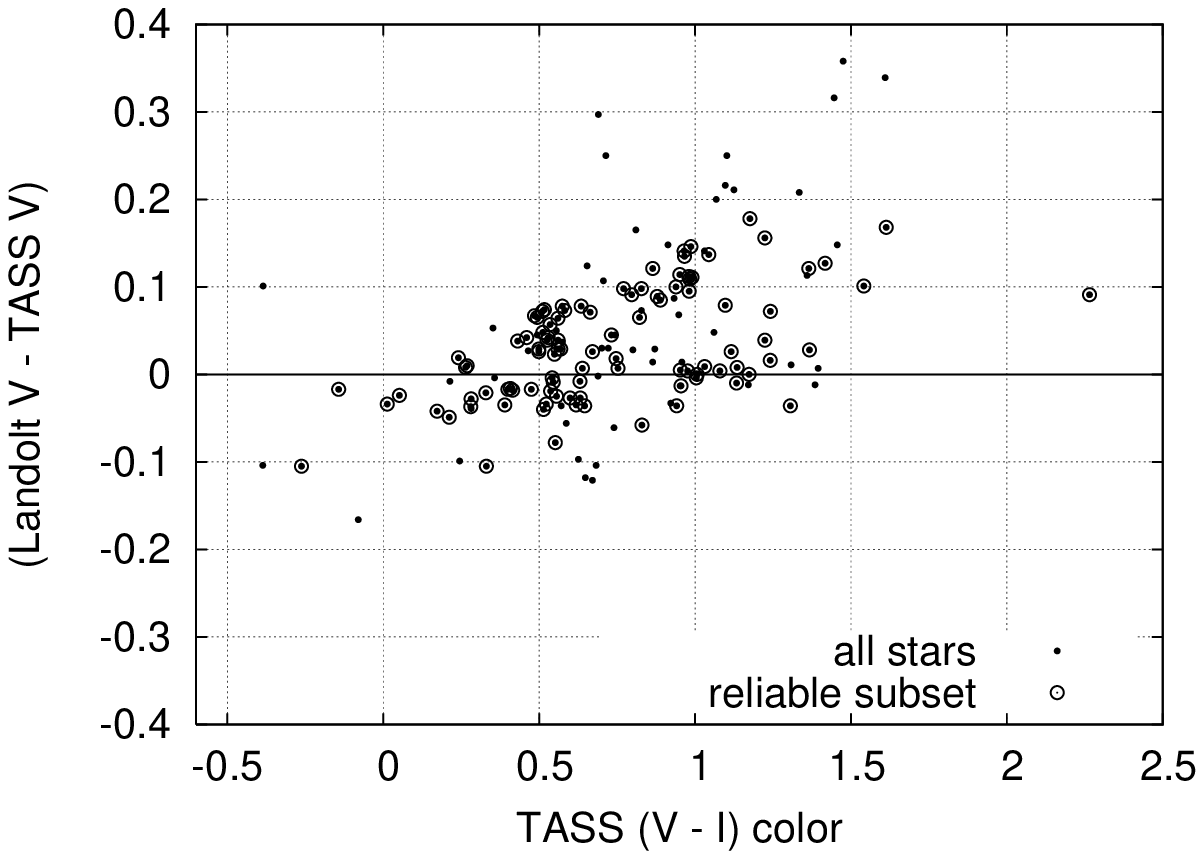}
\caption{Differences in $V$-band between Landolt photometry and Mark IV
           engineering photometry as a function of stellar color
           \label{rawvmagdiffcolor}.}
\end{figure}

\begin{figure}
\plotone{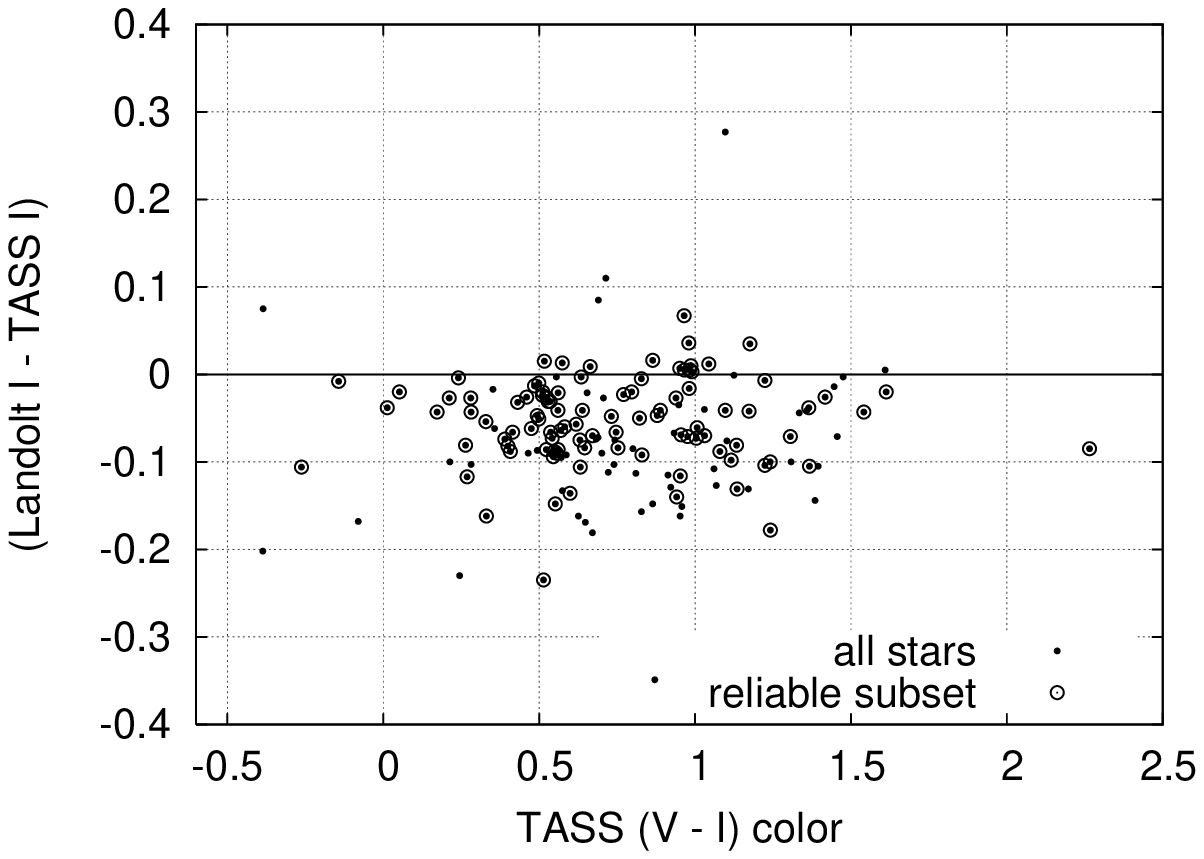}
\caption{Differences in $I_C$-band between Landolt photometry and Mark IV
           engineering photometry as a function of stellar color
           \label{rawimagdiffcolor}.}
\end{figure}

\begin{figure}
\plotone{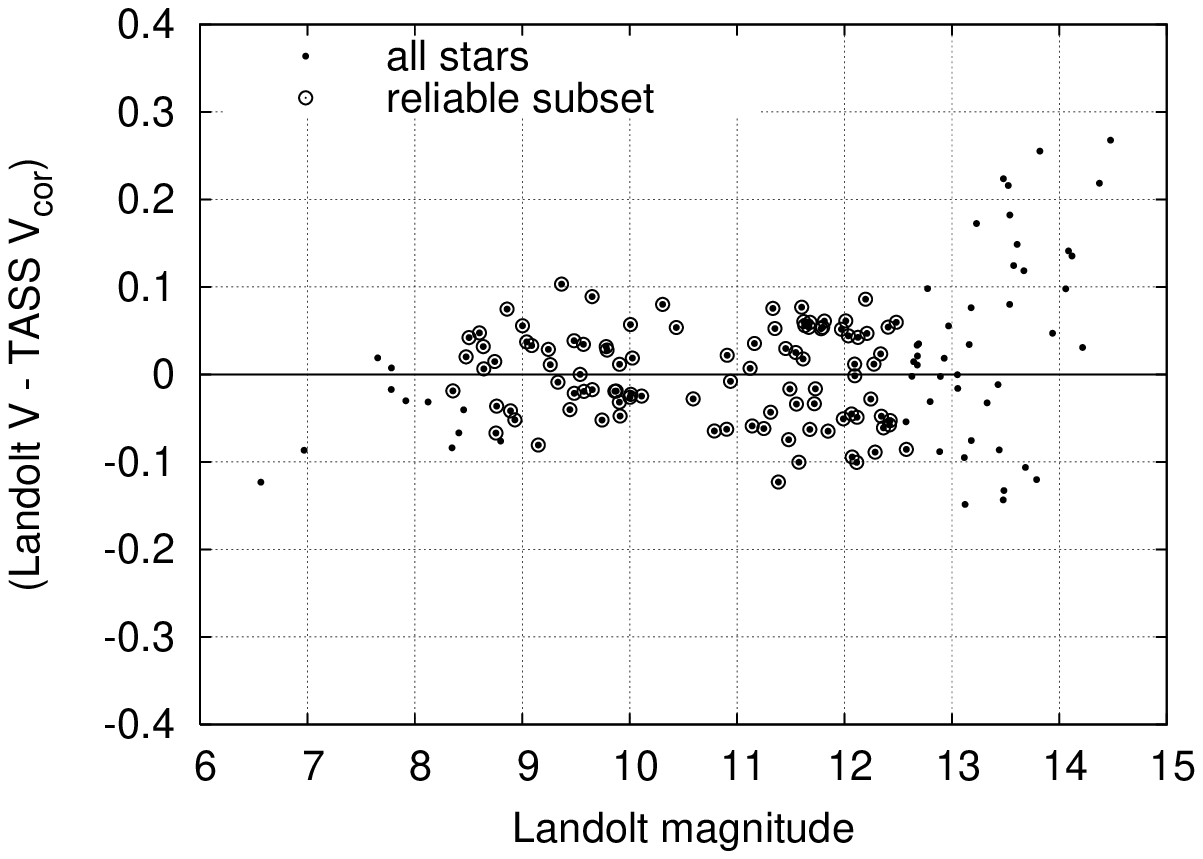}
\caption{Differences in $V$-band between Landolt photometry and 
           corrected Mark IV magnitudes
           \label{corvmagdiff}.}
\end{figure}

\begin{figure}
\plotone{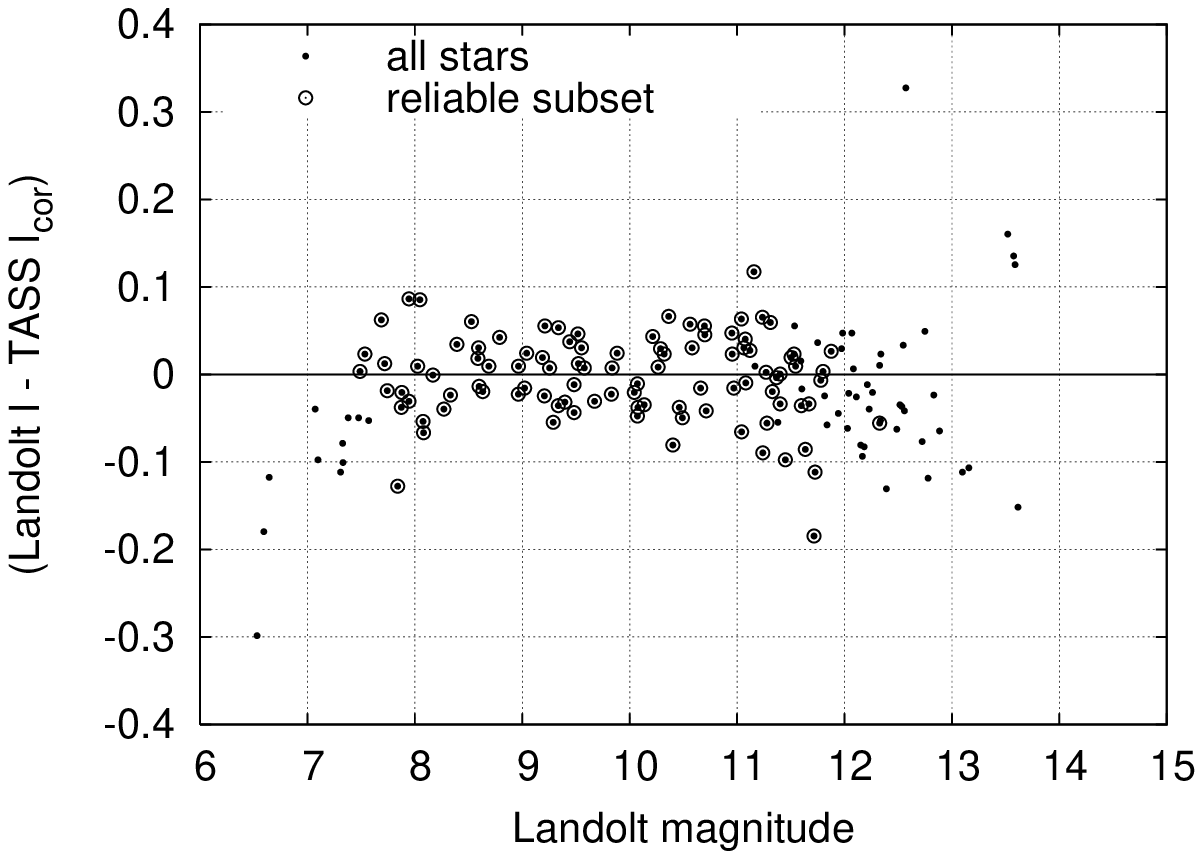}
\caption{Differences in $I_C$-band between Landolt photometry and 
           corrected Mark IV magnitudes
           \label{corimagdiff}.}
\end{figure}

\begin{figure}
\plotone{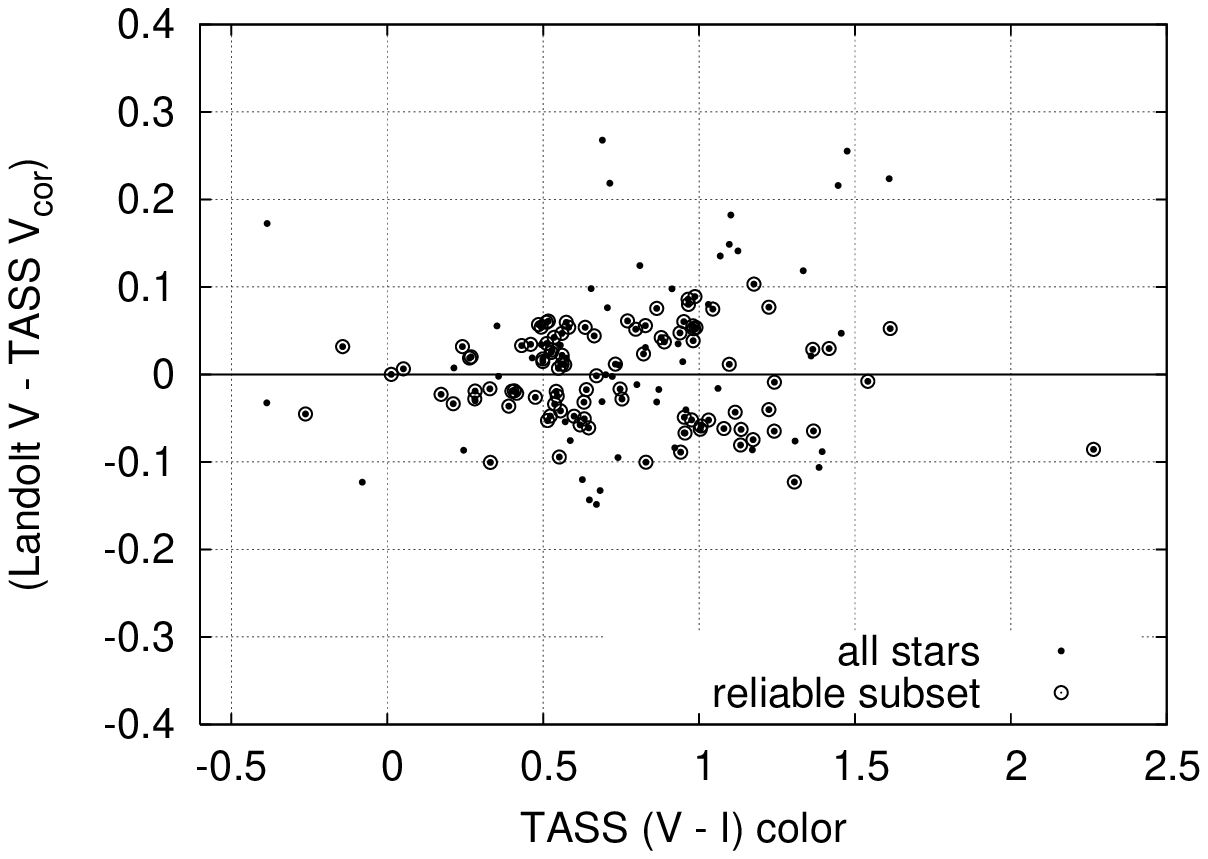}
\caption{Differences in $V$-band between Landolt photometry and 
           corrected Mark IV magnitudes as a function of stellar color
           \label{corvmagdiffcolor}.}
\end{figure}

\begin{figure}
\plotone{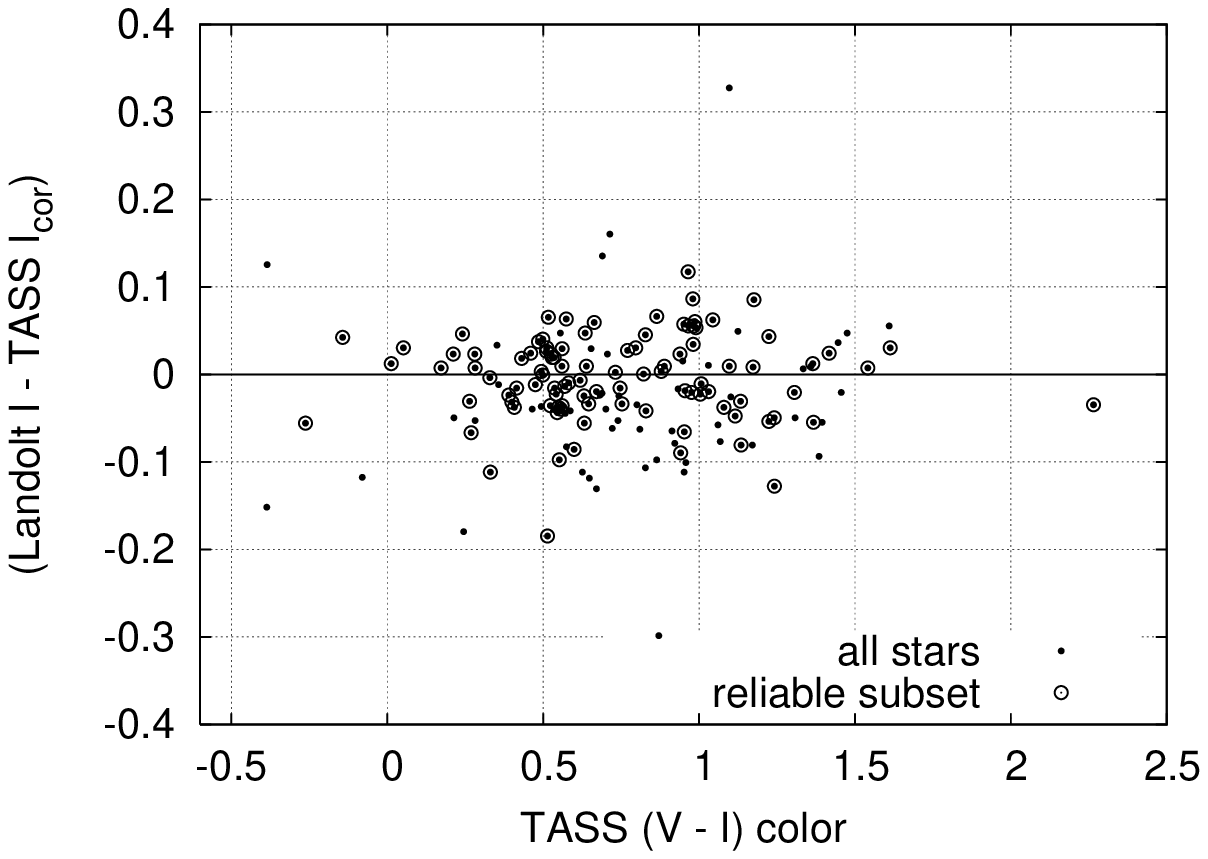}
\caption{Differences in $I_C$-band between Landolt photometry and 
           corrected Mark IV magnitudes as a function of stellar color
           \label{corimagdiffcolor}.}
\end{figure}

The differences between the Landolt magnitudes 
of these stars and the IQM of Mark IV magnitudes
are shown as a function of magnitude in
figures ~\ref{rawvmagdiff}
and
\ref{rawimagdiff},
and as a function of ($V - I_C$) color
in 
figures ~\ref{rawvmagdiffcolor}
and
\ref{rawimagdiffcolor}.
There is clearly a color-dependent error
in the $V$-band measurements, and a fixed offset
in the $I_C$-band measurements.
We made unweighted linear fits to the 
residuals in the reliable subset
to derive the following corrections,

\begin{eqnarray}
V_{\rm cor} &=& V_{\rm eng} - 0.0353 + 0.09362*(V_{\rm eng} - I_{\rm eng}) \\
I_{\rm cor} &=& I_{\rm eng} - 0.0503 
\end{eqnarray}

where $V_{\rm eng}$ and $I_{\rm eng}$ are the interquartile means
of values in the Mark IV engineering database,
and $V_{\rm cor}$ and $I_{\rm cor}$ are the corrected values.
Table~\ref{cormag_landolt_table} 
lists the statistics of the differences
between Landolt photometry and the corrected
Mark IV photometry;
we show the residuals as a function of 
magnitude in 
figure~\ref{corvmagdiff}
and
figure~\ref{corimagdiff},
and as a function of ($V - I_C$) color
in 
figure~\ref{corvmagdiffcolor}
and
figure~\ref{corimagdiffcolor}.

After these corrections, we note that the
errors in Mark IV photometry do not
show the usual pattern of increasing gradually
from bright stars to faint stars;
instead, the errors seem independent of 
magnitude.
Calculation of the errors based on photon
statistics and sensor properties
yield values which are much smaller than the observed
errors for bright stars;
for example, we expect the uncertainty per 
measurement to be less than $0.01$ mag
for stars of $V \leq 10$.
Why do the measurements fail to meet 
these predictions?

One reason is the relatively low precision 
of the preliminary photometric calibration.
As table~\ref{tychotable} indicates,
we accepted Tycho-2 stars with estimated
uncertainties of up to $0.05$ mag
to serve as references in our pipeline
processing;
it was the only way to ensure a reasonable
number of reference stars in every image.
In addition, the particular set of
Tycho-2 stars used to calibrate one particular
target star varied during the first three
years of the survey since the Right Ascension
of image centers drifted randomly from
night to night.
It is possible that this changing mix of 
references could add some noise to the
overall photometric calibration.
However, we note that in 2005, when the 
Right Ascension centers of images were
fixed, and a uniform set of reference
stars did appear repeatedly for each 
target star, the internal scatter of
our measurements did not decrease
significantly.
We do know of 
one additional and significant
source of error:
systematic variations in
photometry as a function of position in the focal
plane, as we will demonstrate in the next section.

\subsection{Using patches to reduce systematic errors}

\begin{figure}
\plotone{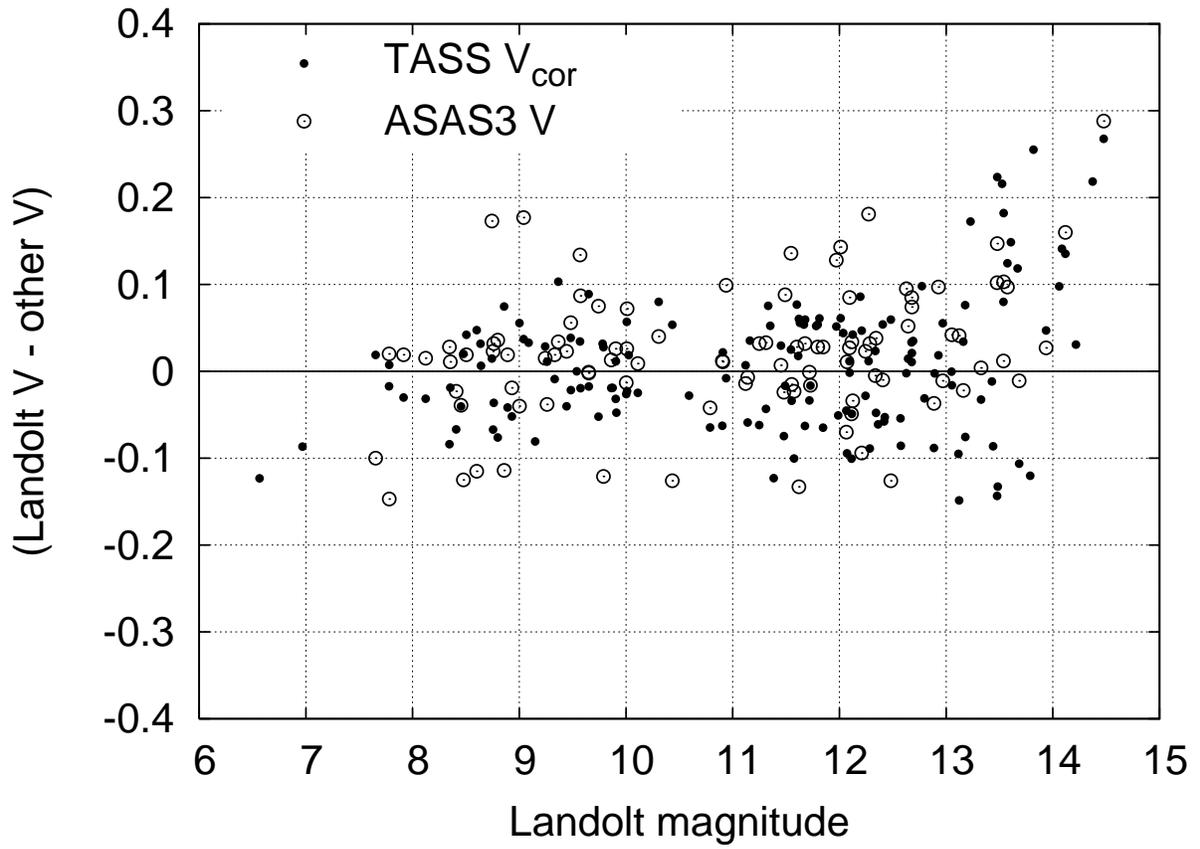}
\caption{Differences from Landolt photometry in $V$-band 
           for Mark IV corrected measurements 
           and ASAS3 measurements \label{comptassasas}.}
\end{figure}

It is not an easy matter to measure precisely the brightness of
astronomical sources over a wide field:
optics deliver a variable illumination
and a variable PSF to the focal plane, 
and the sky brightness varies significantly across the field.
Systematic errors may easily dominate the error budget
for bright sources, as in, for example, 
the Digitized Second Palomar Observatory Sky Survey 
{\citep{Gal2004}},
the ROTSE-I survey
{\citep{Akerlof2000}},
or the All Sky Automated Survey
{\citep{Pojmanski2002}} 
(see figure~\ref{comptassasas}).
If one chooses a fixed set of field centers and points to 
them accurately, then each star may suffer the same error
repeatedly, and thus
the scatter of individual measurements around the mean will be small.
This satisfies the needs of some projects,
such as searching for variable stars;
but for other goals, such as creating a uniform set of
photometric reference stars,
one must remove the errors in mean values 
which remain over the field of view.

The Mark IV survey was carried out so that cameras 
returned to nearly the same Declination each night,
but (for a large fraction of the survey)
varied in Right Ascension in an irregular fashion.
Thus, photometric errors due to location in the focal plane
would appear in a complicated manner.
Consider two identical stars separated by $\sim 2^{\circ}$ 
on the sky
(about half the size of one image)
in the North-South direction;
one might be incorrectly measured as $0.04$ mag 
brighter than the other, night after night after night.
On the other hand, measurements of two identical stars 
separated by $\sim 2^{\circ}$ in the East-West direction 
would depend on the exact pointing of each night's image:
on one night, the eastern star might fall near the image center
and the western star near the edge, causing the eastern star to
appear $0.04$ mag brighter; but the next night, the eastern
star might be near the edge and the western star near the center,
causing the eastern star to appear $0.04$ mag 
{\it fainter.}

Ordinary flatfielding procedures such as ours 
can and do remove variations in sensitivity on
very small spatial scales -- among neighboring pixels --
but fail to correct variations on larger spatial
scale;
indeed, if scattered light enters the optics, 
using a flatfield based on diffuse background light
can {\it introduce} photometric errors in the measurements
of point sources
{\citep{Manfroid1995}.}
The scatter between repeated measurements of bright stars
in our engineering database, roughly $\sim 0.05$ mag,
is half the center-to-edge variation in illumination
from our optics (about $10\%$),
suggesting that our flatfielding procedure may have introduced
much of the error.
Since we assign a single magnitude zero-point to all the
stars in each image during our reductions,
there is no way for our standard processing to 
account for such large-scale, position-dependent variations
in sensitivity.

\begin{figure}
\includegraphics[scale=0.75,angle=-90]{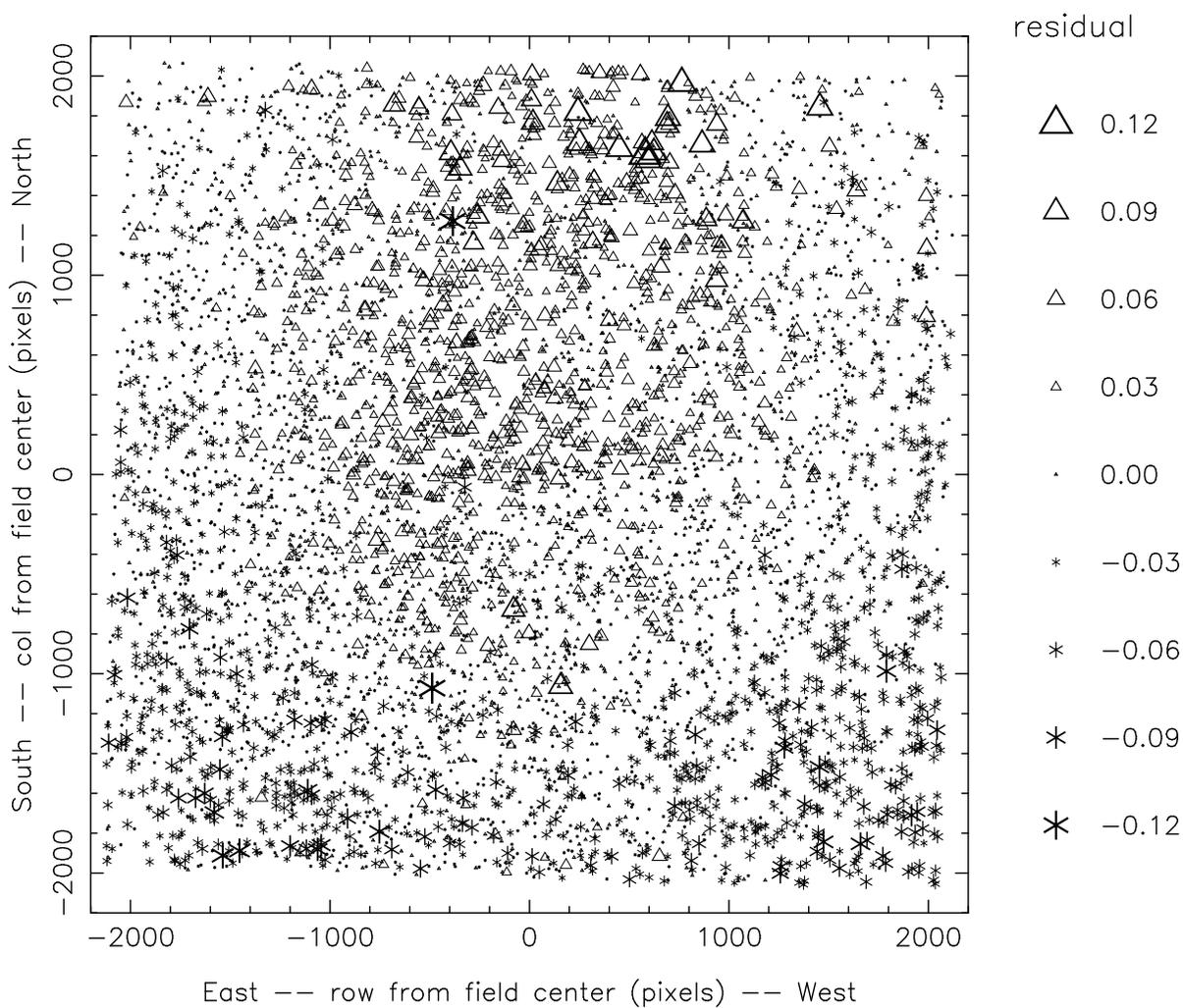}
\caption{Residuals in $V$-band photometry in a special 
           series of ``grid calibration'' images taken by
           the TOM1 unit in November, 2002.  
           The $I_C$-band residuals show a similar pattern.
           \label{residualmapv}}
\end{figure}

As 
{\citet{Manfroid1995}} shows,
it is possible to characterize these large-scale photometric errors
by taking a special set of images arranged in a grid centered
on a rich star field.
In late 2002, we used the TOM1 unit to acquire
images in a $7 \times 7$ grid at four locations 
near the galactic plane.
Our analysis of the grid images 
(see figure~\ref{residualmapv}) 
reveals
a pattern of residuals which has a 
significant change in Declination but 
varies little in Right Ascension,
as one would expect from the fixed-Dec, variable-RA 
positions of fields during much of our survey.
The gradient in Declination is due in large part
to the bright sky at our site;
since we always observe near the meridian,
the bright horizon is always located in the same
direction: to the South, for images taken 
at Declinations $\delta < +42^{\circ}$,
or to the North, for images taken
at Declinations $\delta > +42^{\circ}$.
In the future, we hope to make similar maps of the residuals
in the other two Mark IV units, 
check the maps to verify that they have been stable
over the course of survey, and then apply them to all
the measurements in the engineering database.

Although we cannot at the current time 
correct our photometry for these systematic
errors, 
we can reduce the effect of those errors
on certain statistical properties of our measurements.
As figure~\ref{residualmapv} shows, 
the residuals change gradually across
the $4{\rlap.}^{\circ}2$ field of view.
Suppose that we concentrate on the stars
within a small portion of the field,
one degree on a side.
To first order, all the stars in this
little region will suffer the same
error in measurement each night.
If we perform differential photometry
of the stars in this region alone,
we can recover any variations 
in the light of one star relative to its
neighbors.
The mean magnitudes of stars in the region
will remain uncertain, but we may improve
our knowledge of each star's intrinsic variation
around that mean.

The notion of dividing measurements made across a 
wide field of view into small groups 
is not a new one: 
\citet{Taff1989} 
and
\citet{Bucciarelli1992} 
describe
the benefits of ``imaginary subplates''
for astrometry derived from Schmidt plates.

\begin{figure}
\plotone{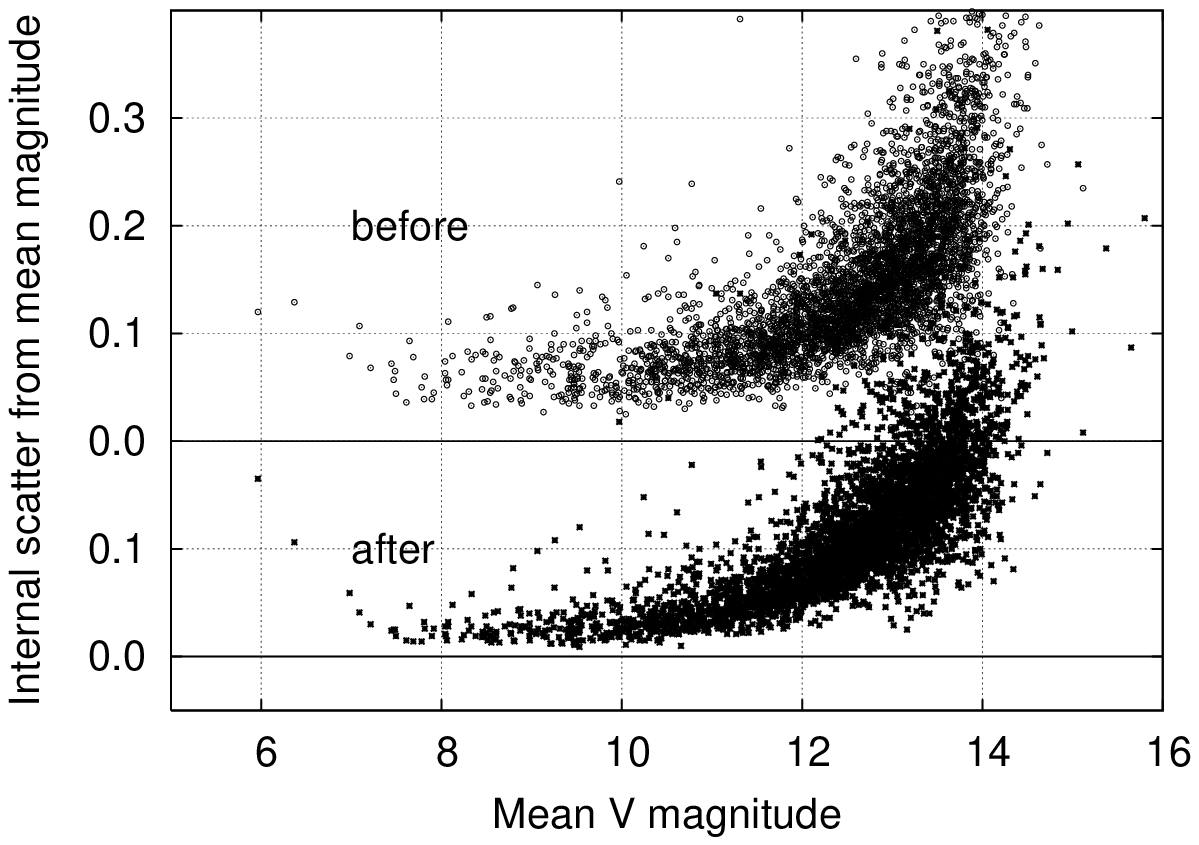}
\caption{Scatter from the mean $V$-band magnitude in repeated measurements,
           before and after ensemble photometry.
           For clarity, only a 
           small random subset of the entire dataset has been 
           plotted.
           \label{sigmavsmagv}}
\end{figure}

\begin{figure}
\plotone{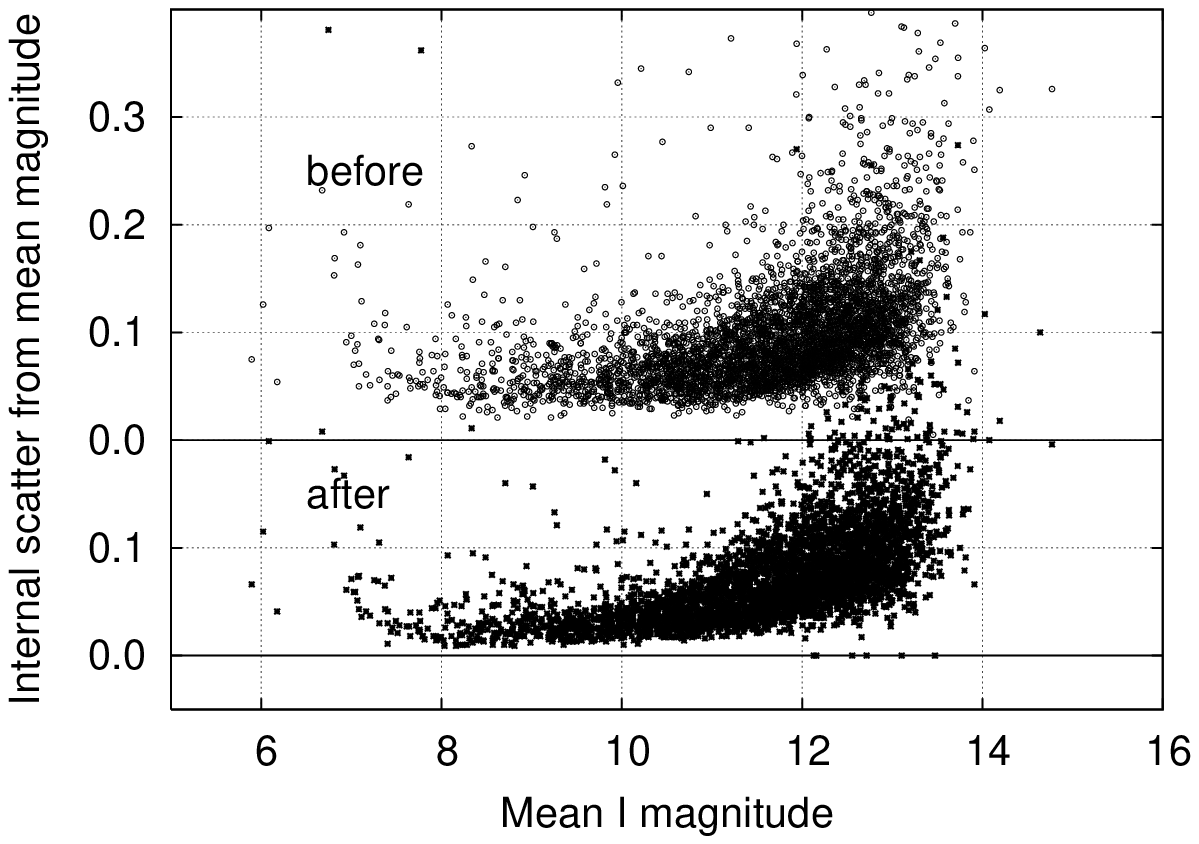}
\caption{Scatter from the mean $I_C$-band magnitude in repeated measurements,
           before and after ensemble photometry.
           For clarity, only a 
           small random subset of the entire dataset has been 
           plotted.
           \label{sigmavsmagi}}
\end{figure}

We divided all the measurements
in our engineering database into ``patches''
one degree on a side;
each patch overlaps its neighbors by one-quarter
of a degree.
The measurements in each patch define
an inhomogeneous ensemble, since faint
stars may not be detected as frequently as bright ones.
Following the methods of 
{\citet{Honeycutt1992}},
we allowed the magnitudes from each image
to shift up or down slightly in order to
minimize the overall differences between 
measurements of the same stars.
The main result for our purposes\footnote{
In theory, the mean ensemble magnitudes 
might replace our IQM magnitudes.
However, the overall zeropoint of each ensemble
is arbitrary, and there are in some star-poor
regions of the sky only 2 or 3 Tycho-2 stars
within each patch to reset the zeropoint;
there could be significant jumps in the zeropoint
from one patch to the next.
Using the overlapping regions of the 
$\sim 40000$ patches to solve for optimal
zeropoints, 
as in 
\citet{Maddox1990},
is an interesting problem beyond
our current abilities.}
is the standard deviation from the mean
ensemble magnitude for each star.
As we show in 
figure~\ref{sigmavsmagv}
and
figure~\ref{sigmavsmagi},
the standard deviation from the mean magnitude
decreases significantly after ensemble
processing:
the floor of the distribution for bright stars
shrinks from $\sim 0.05$ mag to $\sim 0.02$ mag 
in both passbands.

\begin{figure}
\plotone{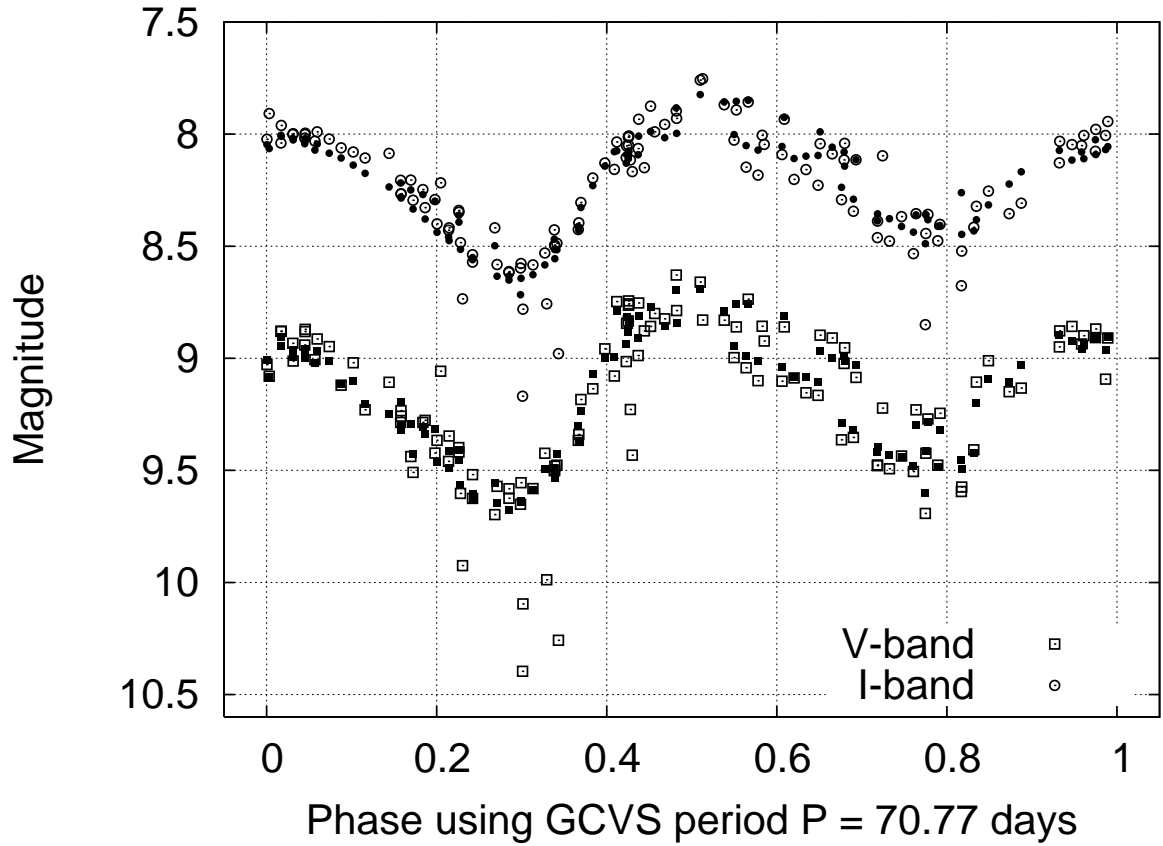}
\caption{Phased light curve of R Sge, showing the difference
           between measurements in the engineering database
           (open symbols) and after ensemble photometry 
           (filled symbols).\label{phasedlightcurve}.}
\end{figure}

Another indication of the improvement
offered by ensemble photometry is visible
in figure~\ref{phasedlightcurve},
which shows the light curve of R Sge
folded according to its period in
the GCVS 
\citep{Samus2004}.
The entire set of measurements in the engineering
database (open symbols) includes several
outliers, and the width of the locus is considerable.
The ensemble procedure (filled symbols) discards data from several
nights deemed to have residuals of larger size
than usual, and brings the remaining measurements
into better agreement.

\subsection{Construction of the ``patches'' catalog
                \label{make_catalog_section}}

The 
``patches'' catalog provides a short
summary of the Mark IV survey:
it contains mean positions and magnitudes
for a subset of the most reliable
detections:
stars seen in a simultaneous ($V$, $I_C$) image
pair on at least five occasions.
We also compute several statistics which indicate
for each star its degree of variability 
in our measurements.
Readers may query the engineering database
for the full (uncorrected, pre-ensemble) photometry of any stars
of interest.

\begin{figure}
\plotone{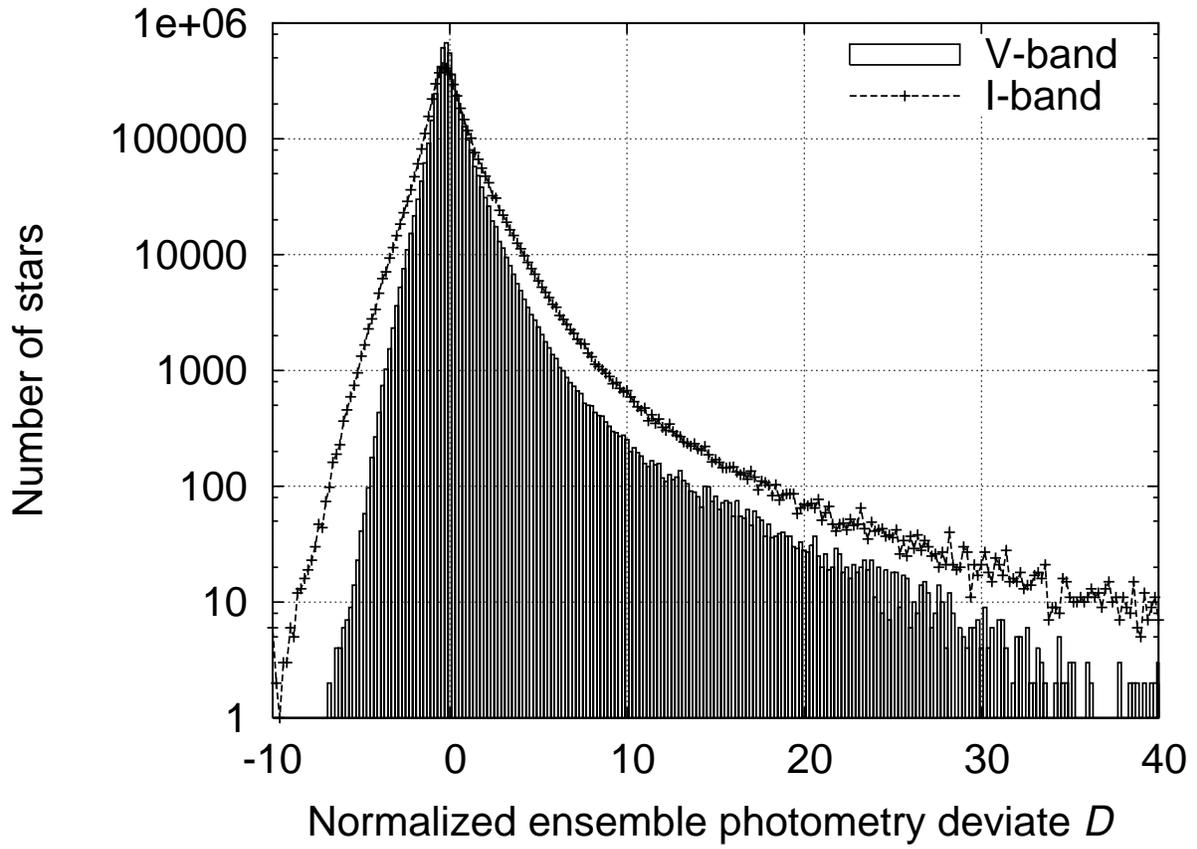}
\caption{Histogram of the normalized deviation from the 
           ensemble median magnitude.  
           \label{deviatehist}}
\end{figure}

We can estimate the variability of a
star in two ways.
First, the ensemble photometry procedure
produces not only a mean differential
magnitude for each star, but also the 
standard deviation of the adjusted measurements
around that mean (henceforth called $\sigma$).
This is the quantity plotted on the ordinate
of 
figure~\ref{sigmavsmagv}
and
figure~\ref{sigmavsmagi}.
We divide the stars into bins by magnitude
and compute the median of 
$\sigma$ 
within each bin,
as well as the range $r$ between the 
first and third quartiles.
We then fit a parabolic model to the median
values as a function of magnitude.
In order to find stars which vary much more
than the typical amount, stars which would lie
far above the main locus in the graph of
$\sigma$ versus magnitude,
we compute a quantity 
we denote $\mathcal{D}$ 
(for its similarity to the normal deviate)
as follows:
\begin{equation}
{\mathcal{D}} = { {{\sigma} - {\sigma_p}} \over {r} }
\end{equation}
where $\sigma$ is the standard deviation of some
star around its ensemble mean,
$\sigma_p$ is the predicted standard deviation
from our model, and $r$ is the average width
of the locus in the $\sigma$ versus magnitude graph.
In essence, $\mathcal{D}$ is a normalized measure of
the degree to which a star varies more than the 
typical star in the ensemble.
Since the ensemble solutions for each passband are
independent, each star is assigned two values of
$\mathcal{D}$.
In 
figure~\ref{deviatehist},
we show the distribution of $\mathcal{D}$ 
in each passband;
the long tails of positive values contain
candidates for variability.

Another measure of variability combines
the information from the two passbands:
the Welch-Stetson 
variability index 
{\citep{Welch1993}}
assumes that changes in a star's luminosity
occur nearly simultaneously at all optical wavelengths.
We use the ensemble output 
to compute
a slightly modified version of the
Welch-Stetson 
statistic
which we shall call $\mathcal{W}$:
\begin{equation}
\mathcal{W} \equiv {\sqrt{ {n}\over{n-1}}} \thinspace \thinspace {\sum_{i=1}^{n}  
          {
            {\left (
                 \left ( {{V_i - \bar{V}}\over{\sigma_V}} \right )
                 \left ( {{I_i - \bar{I}}\over{\sigma_I}} \right ) 
            \right ) } } }
\end{equation}

\begin{figure}
\plotone{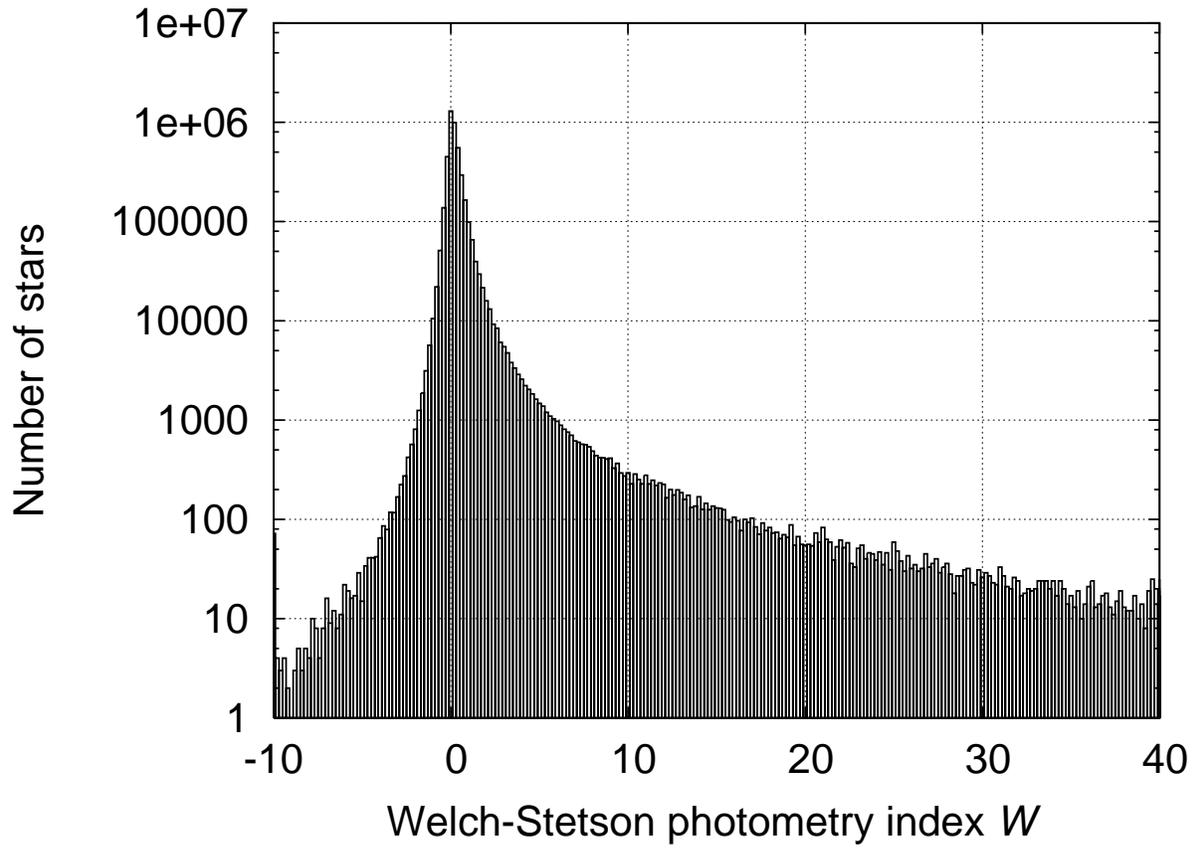}
\caption{Histogram of the Welch-Stetson variability index $\mathcal{W}$.
           \label{welchstetsonhist}}
\end{figure}

Here $V_i$ is the $V$-band ensemble measurement of a star in the $i$-th image,
and $\bar{V}$ the ensemble mean magnitude of the star;
the weighting factor
$\sigma_{V}$ is determined 
from the typical scatter from the ensemble mean
for stars of similar brightness.
The $I_i$, $\bar{I}$ and $\sigma_{I}$ symbols refer to the
analogous quantities in the $I_C$-band ensemble.
The distribution of $\mathcal{W}$ 
(figure~\ref{welchstetsonhist})
shows the same sort of long positive tail
as 
figure~\ref{deviatehist};
some of these outliers are due to intrinsic
stellar variability, others due to erroneous
measurements.

Because the Mark IV images have relatively
large pixels ($7^{''}$ on a side), 
and because we use a large synthetic
aperture 
($30^{''}$ in radius) 
to measure them,
a significant fraction of our measurements 
are contaminated by neighboring stars.
In order to flag stars with possible contamination,
we have looked for nearby companions to each
star, using the survey itself as a reference.
We checked for neighbors within two distances:
$30^{''}$ 
and
$60^{''}$.
We assign a ``proximity code'' to each
star following the rules listed in
table~\ref{proximity}.
For example, a star which has two fainter
neighbors at distances of 
$18^{''}$ 
and
$27^{''}$,
and one brighter neighbor at a distance of
$45^{''}$,
would be assigned a proximity code of 6.
Neighbors which are too close for our equipment 
to resolve, or too faint for our equipment to detect, 
will not be flagged by this procedure.
We find that roughly $0.3\%$ of the stars in our
catalog are marked as having neighbors within 
$30^{''}$ 
and roughly $6.6\%$ marked as having neighbors within 
$60^{''}$.
Our measurements for any of these stars 
may be unreliable.

The final version of our catalog
appears in 
table~\ref{markiv_patches_table}.
It contains 4,353,670 stars
with Declinations in the range
$-5.5 < \delta < +88.2$.

\section{Conclusion}

Using wide-field telescopes and cameras of our own design,
we have measured stars in the rough range
$7 \sim V < 13$
over a four-year period
in the Johnson-Cousins $V$ and $I_C$ passbands.
All of our measurements are freely available
in a database which may be queried over the
Internet.
We have selected a subset of objects observed 
multiple times, 
performed extra photometric calibration,
and computed statistical indications of variability
to create the
``Mark IV patches catalog.''

We believe that this catalog will be especially
useful to 
\begin{itemize}
\item calibrate comparison stars in the fields of
          variable stars, supernovae, and gamma-ray bursts
\item provide a net of photometric comparison stars
          for moving objects 
\item study variable stars of large amplitude
\item verify that certain stars do {\it not}  
          vary above a certain level
\end{itemize}

We remind the reader that our survey does suffer
certain shortcomings:
its large pixels make object detection and measurement
unreliable in even moderately crowded fields,
its photometric measurements contain systematic 
errors at the level of about five percent,
and the engineering database 
contains some data taken in poor conditions.
We chose to acquire very large amounts of information
from a suburban site rather than to scan
small pieces of the sky from 
a clear, dark site,
based in part upon our available resources.

The Mark IV survey continues to collect data:
at the time of writing (September 2006),
the engineering database contained over 190 million
measurements.
We will continue to make our data available 
in several formats
\footnote{
The engineering database 
\url{http://sallman.tass-survey.org/}
for queries,
an archive of flat ASCII files
\url{http://crocus.physics.mcmaster.ca/TASSData/}
for bulk transfers,
and the ``patches catalog'' at SIMBAD
\url{http://simbad.u-strasbg.fr/Simbad}
}
to the community at large for the foreseeable future.


\acknowledgments

The Amateur Sky Survey draws upon the talents
of a group of participants which has averaged about 160
based on the membership of our E-mail list.
A note to the list with a problem often produces
an expert.  For example, during the development of the Mark IV,
we found large and unexpected coma
in our images.  
Two experts came forward and checked the Zemax lens calculations,
which were found to be correct.  
Following their suggestions, we examined one lens element
carefully and discovered that the manufacturer had
mounted it backwards.
We wish to thank particularly
our correspondents
Chris Albertson,
Paul Bartholdi,
Andrew Bennett, 
Robert Creager,
Shawn Dvorak,
Michael Gutzwiller,
Herb Johnson,
Mike Koppleman,
Peter Mount,
Maciej Reszelski,
Jure Skvarc, 
Ron Wickersham,
Patrick Wils,
and Seiichi Yoshida.
Elliot Burke designed the optics for the Mark IV units,
and Dave Garnett was invaluable in their construction.
Many professional astronomers 
contributed as well:
Doug Welch (who hosts a copy of the Mark IV data),
Arne Henden,
and Brian Skiff,
among others.
Bohdan Paczynski
inspired and encouraged
the project from its earliest stages.
This research has made extensive use of the SIMBAD database,
operated at CDS, Strasbourg, France.
MPS would like to thank his wife and kids for their support,
letting him spend far more time 'working' 
on the computer than he probably should.
MWR thanks the RIT Physics Department and College of Science 
for their continued support.
Over 90\% of the expense of parts, construction, fabrication, and
operation has been borne by the first author. This in lieu of traveling
the world or some such in retirement.  Try it, you will like it.


\appendix

\section{The {\bf match} package}

We describe here in some detail the procedure we use to 
calibrate the positions of objects in our images,
since the task is a common one and other astronomers may
be able to adapt our software to meet their needs
(as the SHASSA \citep{Gaustad2001}
and ACS \citep{Blakeslee2003}
teams already have).
Detailed documentation and the full source code can be
found on-line\footnote{
\url{http://spiff.rit.edu/match}}
or by contacting the second author.

The task is to convert 
the pixel coordinates $(x, y)$ of a list of
objects found in some image
to celestial coordinates
$(\alpha, \delta)$.
We assume that the user has a reference catalog of objects
with celestial coordinates
which overlaps substantially (and preferably surrounds)
the detected objects.
We break the job into five steps:
\begin{enumerate}
\item Project the reference objects onto a 
           plane, converting $(\alpha, \delta)$  into
           standard coordinates $(\xi, \eta)$
\item Match the detected objects to the reference objects
\item Find the transformation which takes pixel coordinates
           $(x, y)$ to standard coordinates $(\xi, \eta)$
\item Apply the transformation to all the detected objects
\item De-project the $(\xi, \eta)$ coordinates of the
           detected object onto the celestial sphere,
           yielding their $(\alpha, \delta)$ positions
\end{enumerate}

Our package contains short functions 
to perform the first and last steps,
which involve nothing more than a bit of spherical trigonometry.
The real work lies in the second step:
finding the best set of matches between the detected
and reference objects.
We follow the method described by
\citet{Valdes1995}.
It creates one set of triangles using the detected
objects, a second set of triangles using 
the reference objects,
then searches for similar triangles.
The strength of this technique is its insensitivity
to rotation, translation, inversion and 
differences in scale between the two lists of objects;
its main weakness is that the computing time
required to find a match grows as the total number of
objects raised to sixth power.

Our implementation provides three standalone
programs:
{\tt project\_coords} performs the first step,
{\tt match} steps two through three,
and 
{\tt apply\_match} steps four and five.
In order to make the software flexible
and easily incorporated into existing frameworks
without modifying its source code,
we have given the {\tt match}
program many command-line options,
permitting the user to place contraints
on the number of objects matched, 
the range of relative scale factors of the two lists,
the critical matching radius,
the number of iterations to make, 
etc.
One can also request various amounts of 
diagnostic output
from the routine to verify that a valid
match was found.
The software runs quickly enough to meet
our needs for the Mark IV survey:
it can match two lists of 100 objects
in a few seconds on a typical desktop computer.


\clearpage

\clearpage

\clearpage

\clearpage

\clearpage

\clearpage

\clearpage

\clearpage

\clearpage

\clearpage

\clearpage

\clearpage

\clearpage

\clearpage

\clearpage

\clearpage

\clearpage

\clearpage

\clearpage

\clearpage

\clearpage

\begin{deluxetable}{ll}
\tablecaption{Summary of Mark IV hardware \label{hardware_summary}}
\tablewidth{0pt}
\tablehead{
}
\startdata
CCD chip               &  Loral Fairchild CCD442A        \\
pixels per chip        &  $2048 \times 2048$, each $15 \mu \times 15 \mu$  \\
gain                   &  $2.4 e^{-}$ per ADU \\
readout noise          &  $15 e^{-}$ \\
operating temperature  &  $-20 \thinspace ^{\circ}{\rm C}$ \\
dark current           &  $0.2 e^{-}$ per second per pixel \\
optics                 &  $100{\rm mm \ } f/4$ \\
plate scale            &  $7{\rlap.}^{''}7$ per pixel \\
field of view          &  $4{\rlap.}^{\circ}2 \times 4{\rlap.}^{\circ}2$ \\

\enddata
\end{deluxetable}

\clearpage

\begin{deluxetable}{lcc}
\tablecaption{Subsets of the Tycho-2 catalog \citep{Hog2000}\label{tychotable}}
\tablewidth{0pt}
\tablehead{
\colhead{property} & \colhead{astrometric subset} & \colhead{photometric subset}
}
\startdata
$B_T$ magnitude        &  $1.0 < B_T $   &  $1.0 < B_T < 11.8$         \\
$V_T$ magnitude        &  $1.0 < V_T $   &  $1.0 < V_T < 10.7$   \\
uncertainty in $B_T$   &  $\sigma(B_T) < 0.20$   & $\sigma(B_T) < 0.05$ \\
uncertainty in $V_T$   &  $\sigma(V_T) < 0.20$   & $\sigma(V_T) < 0.05$ \\
color                  &  $-0.2 < (B_T - V_T) < 1.8$ & $-0.2 < (B_T - V_T) < 1.8$ \\
no other Tycho star    &  within $20^{''}$ &  within $50^{''}$           \\
flags                  &  not a double & not a double \\
number of stars        &  1475875   & 360741                 \\

\enddata
\end{deluxetable}

\clearpage

\begin{deluxetable}{cl}
\tablecaption{Converting Tycho-2 magnitudes to the Johnson-Cousins scale\tablenotemark{a}\label{convert_mags}}
\tablewidth{0pt}
\tablehead{
\colhead{Johnson-Cousins} & \colhead{Tycho-2}}
\startdata
$B$                      & $B_T + 0.018 - 0.2580(B_T-V_T) $ \\
$V$                      & $V_T + 0.008 - 0.0988(B_T-V_T) $ \\
$R_C$                    & $V_T - 0.014 - 0.5405(B_T-V_T) $ \\
$I_C$                    & $V_T - 0.039 - 0.9376(B_T-V_T) $ \\

\enddata
\tablenotetext{a}{We use only the $V$ and $I_C$ conversions, but include others for readers with their own needs.}
\end{deluxetable}

\clearpage

\begin{deluxetable}{lcccc}
\tablecaption{Differences between Landolt and corrected Mark IV photometry\label{cormag_landolt_table}}
\tablewidth{0pt}
\tablehead{
\colhead{} & 
       \colhead{N} & 
       \colhead{$(V_{\rm Landolt} - V_{\rm cor})$} & 
       \colhead{N} &
       \colhead{$(I_{\rm Landolt} - I_{\rm cor})$} 
}
\startdata
unclipped & 99  & $-0.016 \pm 0.036$  & 99  & $-0.011 \pm 0.035$  \\
clipped\tablenotemark{a} & 94 & $-0.014 \pm 0.031$  & 92 & $-0.006 \pm 0.024$ 
\enddata
\tablenotetext{a}{After one round of $2 \sigma$ clipping.}
\end{deluxetable}

\clearpage

\begin{deluxetable}{crc}
\tablecaption{How to compute the proximity code for a target star\label{proximity}}
\tablewidth{0pt}
\tablehead{
\colhead{If $\geq 1$ neighbor within } & \colhead{which is} & \colhead{then add}
}
\startdata
$30^{''}$       &   brighter than target &   8 \\
$30^{''}$       &   fainter than target &    4 \\
$60^{''}$       &   brighter than target &   2 \\
$60^{''}$       &   fainter than target &    1
\enddata
\end{deluxetable}

\clearpage


\begin{deluxetable}{rrrrrrrrrrrrrrrr}
\tabletypesize{\scriptsize}
\rotate
\tablecaption{Mark IV patches catalog\label{markiv_patches_table}}
\tablewidth{0pt}
\tablehead{
       \colhead{TASS ID} & 
       \colhead{N\tablenotemark{a}} & 
       \colhead{RA\tablenotemark{b}} & 
       \colhead{$\sigma({\rm RA})$} & 
       \colhead{Dec\tablenotemark{b}} & 
       \colhead{$\sigma({\rm Dec})$} & 
       \colhead{V\tablenotemark{c}} & 
       \colhead{$\sigma(V)$\tablenotemark{d} } & 
       \colhead{V ens $\sigma$\tablenotemark{e}} & 
       \colhead{I\tablenotemark{c}} & 
       \colhead{$\sigma(I)$\tablenotemark{d} } & 
       \colhead{I ens $\sigma$\tablenotemark{e}} & 
       \colhead{$\mathcal{D}_V$\tablenotemark{f}} & 
       \colhead{$\mathcal{D}_I$\tablenotemark{f}} & 
       \colhead{$\mathcal{W}$\tablenotemark{g}} & 
       \colhead{prox\tablenotemark{h}} 
}
\startdata
  397943 &  58 &  0.00042 & 0.00018  & -5.49426 &  0.00016 &  9.285 & 0.088 & 0.041 &  8.194 & 0.030 & 0.022  &  0.21 &  0.45 &  -0.07 & 0 \\
  398770 &  23 &  0.00061 & 0.00063  &  5.38051 &  0.00062 & 13.246 & 0.158 & 0.138 & 12.025 & 0.099 & 0.072  & -0.26 & -0.02 &   0.40 & 0 \\
  366763 &  64 &  0.00085 & 0.00012  &  1.08900 &  0.00008 &  9.095 & 0.044 & 0.027 &  8.561 & 0.023 & 0.022  & -0.59 & -0.42 &   0.57 & 0 \\
  398771 &  30 &  0.00116 & 0.00050  &  3.02078 &  0.00070 & 13.030 & 0.133 & 0.112 & 12.162 & 0.065 & 0.055  & -0.46 & -1.31 &   0.22 & 0 \\

\enddata
\tablecomments{Table \ref{markiv_patches_table} is published in 
                its entirety in the electronic edition of the 
                {\it PASP}.  A portion is
                shown here for guidance regarding its form and content.}
\tablenotetext{a}{Number of measurements in the output of ensemble analysis.}
\tablenotetext{b}{Mean of positions in the engineering database, 
                    decimal degrees in equinox J2000.}
\tablenotetext{c}{Interquartile mean of 
                    measurements in engineering database,
                    corrected for color terms.}
\tablenotetext{d}{Standard deviation from the mean of all
                    measurements in engineering database.}
\tablenotetext{e}{Standard deviation from mean magnitude
                    in ensemble photometry of one patch.}
\tablenotetext{f}{Normalized deviation above typical scatter
                    in ensemble solution.}
\tablenotetext{g}{Welch-Stetson variability index.}
\tablenotetext{h}{Proximity code; see section~\ref{make_catalog_section}.}
\end{deluxetable}


\begin{thebibliography}{}
\bibitem[Akerlof et al.(2000)]{Akerlof2000} Akerlof, C., et al. 2000,
                        \aj, 119, 1901
\bibitem[Bakos et al.(2004)]{Bakos2004} Bakos, G., et al. 2004,
                        \pasp, 116, 266
\bibitem[Blakeslee et al.(2003)]{Blakeslee2003} Blakeslee, J. P., et al. 2003,
                        ASP Conf. Ser. 295, 
                        Astronomical Data Analysis Software and Systems XII,
                        ed. H. E. Payne, R. I. Jedrzejewski \& R. N. Hook,
                        (San Francisco: ASP), 257
\bibitem[Bessell(1990)]{Bessell1990} Bessell, M. S. 1990, \pasp, 102, 1181
\bibitem[Bucciarelli et al.(1992)]{Bucciarelli1992} Bucciarelli, B., et al. 
                        1992, \aj, 103, 1689
\bibitem[Cudworth(1985)]{Cudworth1985} Cudworth, K. M. 1985, \aj, 90, 65
\bibitem[Gal et al.(2004)]{Gal2004} Gal, R. R., et al. 2004,
                        \aj, 128, 3082
\bibitem[Gaustad et al.(2001)]{Gaustad2001} Gaustad, J. E., et al. 2001,
                        \pasp, 113, 1326
\bibitem[Henden(2001)]{Henden2001} Henden, A. 2001, private communication
\bibitem[H{\o}g et al.(2000)]{Hog2000} H{\o}g, E., et al. 2000, \aap, 355, L27
\bibitem[Honeycutt(1992)]{Honeycutt1992} Honeycutt, R. K. 1992, \pasp, 104, 435
\bibitem[Howell(1989)]{Howell1989} Howell, S. B. 1989, \pasp, 101, 616
\bibitem[Landolt(1983)]{Landolt1983} Landolt, A. U. 1983, \aj, 88, 439
\bibitem[Landolt(1992)]{Landolt1992} Landolt, A. U. 1992, \aj, 104, 340
\bibitem[Maddox, Efstathiou \& Sutherland(1990)]{Maddox1990} 
                        Maddox, S. J., Efstathiou, G., \& Sutherland, W. J.
                        1990, \mnras, 246, 433
\bibitem[Manfroid(1995)]{Manfroid1995} Manfroid, J. 1995, \aaps, 113, 587
\bibitem[Monet et al.(2003)]{Monet2003} Monet, D. G., et al. 2003, \aj, 125, 984
\bibitem[Paczy\'{n}ski(2000)]{Paczynski2000} Paczy\'{n}ski, B. 2000, 
                        \pasp, 112, 1281
\bibitem[Pojma\'{n}ski(2002)]{Pojmanski2002} Pojma\'{n}ski, G. 2002, 
                        Acta Astronomica, 52, 397
\bibitem[Richmond et al.(2000)]{Richmond2000} Richmond, M. W. et al. 2000, 
                        \pasp, 112, 397
\bibitem[Samus et al.(2004)]{Samus2004} Samus, N. N., et al. 2004, 
            \url{http://www.sai.msu.su/groups/cluster/gcvs/gcvs/}
\bibitem[Stetson(1987)]{Stetson1987} Stetson, P. B. 1987, \pasp, 99, 191
\bibitem[Taff(1989)]{Taff1989} Taff, L. G. 1989, \aj, 98, 1912
\bibitem[Treffers \& Richmond(1989)]{Treffers1989} 
              Treffers, R. R., \& Richmond, M. W. 1989, \pasp, 101, 725
\bibitem[Valdes et al.(1995)]{Valdes1995} Valdes, F. G., et al. 1995, 
              \pasp, 107, 1119
\bibitem[Welch \& Stetson(1993)]{Welch1993} 
              Welch, D. L., \& Stetson, P. B. 1993, \aj, 105, 1813
\end{thebibliography}
\end{document}